\documentclass[12pt,english]{article}
\usepackage{CJKutf8}
\usepackage[utf8]{luainputenc}
\usepackage{babel}
\usepackage{array}
\usepackage{multirow}
\usepackage{varwidth}
\usepackage{amsmath}
\usepackage{amsthm}
\usepackage{amssymb}
\usepackage{stmaryrd}
\usepackage{graphicx}
\usepackage{setspace}
\usepackage[authoryear]{natbib}
\usepackage[bookmarks=false,
 breaklinks=false,pdfborder={0 0 1},backref=section,colorlinks=false]
 {hyperref}
\hypersetup{
 colorlinks,citecolor=blue,pdftex}

\makeatletter

\providecommand{\tabularnewline}{\\}
\usepackage{amsfonts}
\usepackage{amsthm}
\usepackage{lscape}
\usepackage{appendix}
\usepackage{dcolumn}
\usepackage{rotating}
\usepackage{comment}
\usepackage{color}
\usepackage{graphicx}
\usepackage{booktabs}
\usepackage{lmodern}
\usepackage{adjustbox}
\usepackage{caption}
\usepackage{multirow}

\setcitestyle{authoryear,round}

\usepackage{pgf}
\usepackage{tikz}
\usepackage{caption}
\usepackage{subcaption}
\usepackage{tkz-tab}
\usepackage{tikz-3dplot}
\usetikzlibrary{calc}
\usetikzlibrary{arrows, automata}

\newcolumntype{d}[1]{D{.}{.}{#1}}
\newcolumntype{t}[1]{D{,}{,}{#1}}
\newcolumntype{i}[1]{D{.}{}{#1}}

\theoremstyle{definition}
\newtheorem{theorem}{Theorem}[section]

\newtheorem{corollary}{Corollary}[section]

\newtheorem*{example}{Example}

\newtheorem{proposition}{Proposition}[section]

\newtheorem*{asSMAR}{Assumption SMAR}
\newtheorem*{asOVERLAP}{Assumption Overlap}

\newtheorem*{asR}{Assumption R}

\newtheorem*{asM}{Assumption M}

\newtheorem*{example1}{Example (Missing treatment and outcome)}

\numberwithin{equation}{section}

\tikzstyle{level 1}=[level distance=5cm, sibling distance=3.5cm]

\tikzstyle{level 2}=[level distance=5cm, sibling distance=2cm] 

\tikzstyle{bag} = [text width=10em, text centered] \tikzstyle{end} = [circle, minimum width=3pt,fill, inner sep=0pt] 

\tikzstyle{level 1}=[level distance=5cm, sibling distance=3.5cm] 

\tikzstyle{level 2}=[level distance=5cm, sibling distance=2.5cm]

\tikzstyle{bag} = [text width=3cm, text centered]

\tikzstyle{end} = [circle, minimum width=3pt,fill, inner sep=0pt]

\usepackage{pifont}

\newcommand{\E}{\mathrm{E}}
\DeclareMathOperator{\Var}{Var}
\date{}

\makeatother

\begin{document}
\begin{CJK}{UTF8}{}%

\title{A Doubly Robust GMM Estimator for Sequential Non-monotone Missingness}
\author{Shenshen Yang$^{a}$\thanks{Corresponding author. Email: \href{mailto:shenshenyang@tju.edu.cn}{shenshenyang@tju.edu.cn}. I am very grateful to Jason Abrevaya, Sukjin Han, Brendan Kline, Haiqing
Xu, Stephen Donald, Daniel Ackerberg, Dean Spears,
Isaiah Andrews, James Brand, Shaofei Jiang, Xue Li,
Jiangang Zeng, and all participants at the seminars of Peking University,
Zhejiang University and University of Manchester for their helpful
comments and suggestions.} \\[0.3em]
\small $^{a}$Ma Yinchu School of Economics, Tianjin University, \\
\small 92nd Weijin Rd, Building 25, Room 1114, Tianjin 300072, China}
\maketitle
\begin{abstract}
We study moment-based estimation with two sequentially collected variables
subject to non-monotone missingness. The commonly used Missing at
Random (MAR) assumption requiring all missingness mechanisms to depend
on the same fully observed covariates often fails in such cases. We
introduce a sequential MAR assumption that allows asymmetric missingness
mechanisms across stages. Based on this assumption, we construct an
Augmented Inverse-Probability-Weighted GMM (AIPW-GMM) estimator. The
estimator features an asymmetric structure for the augmentation term,
guarantees double robustness, and achieves the closed-form semiparametric
efficiency bound. An application to two-period survey data from the
Oregon Health Insurance Experiment supports the observable implications
of the new assumption. The proposed approach reduces the standard
errors by more than 50\% for the estimated effects of the Oregon Health
Plan among older adults, ``driving'' previously statistically insignificant
estimates significant.

\vspace{0.1in}

%\noindent\textit{JEL Numbers:} C14, C21, C51, I13 

\noindent\textit{Keywords:} Non-monotone missingness; AIPW; GMM;
double robustness.
\end{abstract}

\section{Introduction\protect\label{sec:Introduction}}

This paper considers missingness across two stages. When more than
one variable is missing, the missingness pattern can be classified
as monotone or (strictly) non-monotone. Monotone missingness occurs
when the absence of the first-stage variable implies the absence of
the second-stage variable. In contrast, a non-monotone missing pattern
allows any variable to be observed even when the other is missing.
Let the missing variables be $X_{1}$ and $X_{2}$, and let $R_{1},R_{2}$
denote binary indicators of their observability, where $R_{1}=1$
if $X_{1}$ is observed (and analogously for $R_{2}$). Monotone missingness
imposes $(1-R_{1})R_{2}=0$ almost surely, whereas a non-monotone
missing pattern allows positive probability mass on all combinations
of $(R_{1},R_{2})\in\{0,1\}^{2}$. 

In practice, monotone missingness frequently arises from subsampling
strategies or attrition. However, in most cases, the missing pattern
still exhibits non-monotone features even when data are collected
sequentially. In such cases, incompleteness is not necessarily by
design and can result from self-selection. One common example is longitudinal
survey data, where multistage missingness occurs because participants
skip certain waves or questions. The presence of such a non-monotone
missing pattern introduces additional identification and estimation
challenges. A prevalent approach to handle this issue is to drop all
observations with incomplete data, known as Complete Case (CC) analysis.
This typically yields inefficient estimators due to information loss,
and more importantly, when the missingness mechanism is correlated
with endogenous variables, the CC estimator becomes biased (\citet{little2002bayes};
\citet{qi2014missing}). Therefore, the unbiasedness of a CC estimator
can only be guaranteed under the Missing Completely at Random (MCAR)
assumption, under which the missingness mechanisms are assumed to
be completely random. 

A weaker assumption is Missing at Random (MAR). In the single-missing-variable
case, it requires the missingness mechanism to be independent of the
missing value conditional on observed variables.\footnote{MCAR and MAR can be interpreted as strong ignorability and conditional
ignorability, respectively.} However, with multiple missing variables, partially observed variables
complicate this condition. We typically need to exclude dependence
of the missingness mechanism on partially observed variables to avoid
failure of identification and estimation. This leads to a strengthened
version of MAR assumption in the literature on non-monotone missingness
(\citet{chaudhuri2016gmm}), under which missingness mechanisms for
different variables are independent of their missing values given
the same set of fully observed variables. This assumption also guarantees
identification and yields a closed-form efficient influence function,
from which an Augmented Inverse-Probability-Weighted (AIPW) estimator
can be constructed. 

Despite the usefulness of this assumption in many contexts, such as
models with multiple missing IVs (\citet{chaudhuri2016gmm}; \citet{feng2016instrumental}),
this strengthened MAR assumption fails in many settings where multiple
missing variables are collected sequentially. When variables are missing
simultaneously, their missingness mechanisms are likely to be symmetric.
However, when variables are recorded at different stages, their missingness
mechanisms may differ across stages, and later-stage missingness may
depend on partially observed variables from the previous stage. This
resembles the ``updating feature'' in the monotone missingness literature
(\citet{chaudhuri2020efficiency}), but in the present case, some
observations have $X_{1}$ missing while $X_{2}$ is observed. The
Oregon Health Insurance Experiment (OHIE) survey data present such
a pattern. The survey was conducted in two stages, and we use enrollment
status in the Oregon Health Plan as the first-stage partially missing
variable $X_{1}$, and several health outcome variables as $X_{2}$.
These data provide evidence of a strong correlation between partially
observed $X_{1}$ and $R_{2}$, for each outcome variable. Such patterns
also appear in other multi-period survey data (\citet{dupas2009matters,callen2019headwaters,johnson2025effects}).
Such dependence breaks the strengthened MAR assumption used in previous
studies.

In this paper, we provide a new MAR-type assumption tailored to multistage
settings with non-monotone missingness. This new assumption allows
the later-stage missingness mechanism to depend on partially observed
variables in the previous stage, as presented in the OHIE example.
To avoid identification problems, we allow only the later-stage missingness
mechanism to depend on earlier partially observed variables. We exploit
the sequential feature of the data collection process to justify this
asymmetric missingness assumption, and provide evidence from the OHIE
example to support it. Under a mild additive separability condition
on the moment function, this new assumption still delivers a closed-form
efficient influence function. We then propose a moment function that
retains the AIPW form but includes asymmetric augmentation terms across
stages. The resulting AIPW estimator maintains double robustness and
coincides with the efficient influence function, which is typically
difficult to achieve under non-monotone missingness, especially when
missingness mechanisms differ across stages. In the analysis using
a subsample from the OHIE data, the proposed estimator effectively
increases efficiency and remains unbiased. It reduces standard errors
by more than 50\%, and turns many statistically insignificant estimates
into significant ones at the 0.1\% level. This estimator also attains
the semiparametric efficiency bound when the strengthened MAR assumption
holds, as long as missingness mechanisms are correctly specified.

This paper contributes to the literature in the following aspects:
First, it contributes to the literature on missing data with non-monotone
patterns. Existing studies provide approaches for handling simultaneously
missing variables with non-monotone patterns, as well as multistage
missingness with monotone patterns. However, the case of sequentially
collected data with non-monotone missingness, despite being frequently
encountered in practice, has received little attention. This paper
proposes a MAR-type assumption and a corresponding AIPW estimator
for this specific case and thereby fills the gap. Our framework allows
later-stage variables to be observed even when earlier-stage variables
are missing and permits the later-stage missingness mechanism to vary
depending on the observability of the first-stage variable. This one-way
dependence provides a more flexible and realistic assumption for many
observational datasets, particularly longitudinal surveys.

Second, it is well known that under non-monotone missingness, achieving
double robustness and a closed-form efficient influence function is
challenging. \citet{chaudhuri2016gmm} accomplished this under a new
MAR assumption. They showed the key condition for obtaining a closed-form
efficient influence function and efficiency bound is that all missingness
mechanisms are independent of any missing variable given the same
set of fully observed variables. However, in our setting, this condition
no longer holds. We propose an estimator that retains these desirable
properties even when the key condition fails. This estimator performs
better than those derived from CC analysis and IPW moment conditions,
and its performance surpasses that of the AIPW estimator designed
for monotone missingness, underscoring the importance of considering
the non-monotone component of the dataset. Moreover, under strengthened
MAR with correctly specified missingness propensities, our estimator
continues to maintain these properties and performs comparably to
the estimator of \citet{chaudhuri2016gmm}, indicating the usefulness
of this approach in a broader class of settings.

Third, we show the practical relevance of this approach by justifying
the new assumption and illustrating the AIPW approach with a widely
used dataset in health economics -- the OHIE data. This dataset is
used as a running example in this paper to demonstrate missingness
pattern and support the new assumption. Through simple regression
analyses, we observe a significant correlation between the first-stage
missing variable and later-stage missingness mechanisms, suggesting
that the MAR assumption previously used in the literature is likely
invalid. Using a proxy variable from administrative data, we further
document asymmetry in the later-stage missingness mechanism, depending
on whether the first-stage variable is observed. Focusing on the effects
of the Oregon Health Plan (OHP) for individuals aged 60 and above,
we find that dropping incomplete observations results in insignificant
estimates due to information loss. In contrast, the AIPW approach
yields coefficients very close to the CC estimates, but they are significant
at the 0.1\% level, indicating robust and meaningful effects of Medicaid
on health-related outcomes among older adults. This newly proposed
approach is efficient and easy to implement, following a standard
GMM procedure. 

The rest of this paper is organized as follows. Section \ref{sec:Model-and-Missing-Pattern}
presents the model and missingness patterns; Section \ref{subsec:A-Sequential-Missing}
introduces the key identifying assumption and provides justification;
Section \ref{sec:The-AIPW-GMM-Estimator} proposes the AIPW-GMM estimator
based on this assumption and discusses its statistical properties;
Section \ref{sec:Simulation} illustrates the performance of the estimator
through Monte Carlo simulations; Section \ref{sec:Application-on-the}
applies the proposed method to the OHIE data. The Appendix includes
some technical details and all proofs.

\section{Model and Missing Pattern\protect\label{sec:Model-and-Missing-Pattern}}

We consider a simple two-stage model. Let $X\equiv(X_{1},X_{2})$
denote the incomplete variables collected at successive stages and
let $W$ denote the vector of fully observed variables. The parameter
$\beta^{0}$ is defined as the unique solution to the following moment
condition:
\begin{align}
\beta & =\beta^{0}\text{ if and only if }\E\left[g(X,W;\beta)\right]=0,\nonumber \\
\text{where } & g(X,W;\beta)\equiv g_{1}(X_{1},W;\beta)+g_{2}(X_{2},W;\beta).\label{eq:=000020model}
\end{align}
The individual subscripts are suppressed for simplicity. Let $d_{g}$
denote the dimension of moment conditions and $d_{\beta}$ the dimension
of $\beta$. We allow $d_{g}\geq d_{\beta}$: $d_{g}=d_{\beta}$ corresponds
to the just-identification case, whereas $d_{g}>d_{\beta}$ represents
over-identification. Our primary objective is to identify and consistently
estimate the parameter $\beta$. 

Additive separability is not necessary for identifying $\beta$. However,
it is crucial for later constructing asymmetric augmentation terms
for $X_{1}$ and $X_{2}$ within the AIPW moment function, which ensures
the desired properties of the estimator (discussed later). Such additive
separability arises in many commonly used models. A classical example
is the orthogonal moment conditions in partially linear models, such
as Robinson's partially linear regression and its IV extensions for
potentially endogenous regressors (\citet{newey1994series}; \citet{schick1996efficient};
\citet{allen2019identification}). Other examples include the two-cohort
staggered difference-in-differences design (\citet{sun2021estimating}),
and the partially linear additive spatial autoregressive models (\citet{lu2024gmm}).
Additive separability also characterizes weighted sums of moment conditions,
such as settings with multiple IVs (\citet{newey1990efficient}),
and in system-GMM estimators that stack difference and level moments
(\citet{arellano1995another}).

\begin{example1}

One common missing scenario involves both missing treatment and outcome
variables. Their relationship often satisfies a partially linear structure.
Also, there is typically a natural time gap between the collection
of these two variables to allow treatment to take effect. We retain
the notation $X_{1}$ and $X_{2}$ for treatment and outcome variables,
to emphasize the sequential structure. Assume the simple specification
with additive error $X_{2}=f\left(X_{1};\beta\right)+\epsilon$. Let
$W$ denote fully observed instrumental and other exogenous variables.
Then, the moment function can be written as 
\[
g(X,W;\beta)\equiv W\left(X_{2}-f\left(X_{1};\beta\right)\right)=\underbrace{-Wf(X_{1};\beta)}_{g_{1}}+\underbrace{WX_{2}}_{g_{2}},
\]
which is additively separable. Accordingly, we use missing treatment
and outcome variables as our main running example; the OHIE data provide
an empirical instance of this structure.

\end{example1}

We use $R_{1}$ and $R_{2}$ to denote the observability of $X_{1}$
and $X_{2}$, respectively, formally defined by
\begin{equation}
R_{1}=\begin{cases}
1 & X_{1}\text{ is observed}\\
0 & X_{1}\text{ is missing}
\end{cases},\;\;R_{2}=\begin{cases}
1 & X_{2}\text{ is observed}\\
0 & X_{2}\text{ is missing}
\end{cases}.\label{eq:missing=000020X_1=000020indicator}
\end{equation}

We consider a setting in which the two partially missing variables
are collected in two stages. Leading examples include missing treatment
and outcome variables, where outcomes are typically collected in later
periods for the treatment to take effect; missing instrument variables
and missing treatment; and missing variables in a dynamic setting.
These cases are most frequently discussed in a monotone missingness
setting because of the sequential data structure. However, when the
missingness is not by design, the missingness pattern may still be
non-monotone. Such pattern encompasses a broader spectrum of missing
data patterns and allows $X_{2}$ to be observed even when $X_{1}$
is missing.\footnote{In some literature, non-monotone missingness refers to a more general
missing pattern that includes both monotone and strictly non-monotone
as well as univariate missing patterns (\citet{van2018flexible}).
Other studies, however, focus specifically on the strictly non-monotone
case (\citet{chaudhuri2016gmm}), and we adopt the narrower definition.} 

\subsection{Running Example: OHIE\protect\label{subsec:Example:-Oregon-Health}}

The OHIE survey data are the primary example we use throughout this
paper. In this subsection, we first introduce the background and present
the missing pattern to motivate the subsequent analysis. 

In 2008, a group of low-income individuals was randomly selected by
lottery for the opportunity to apply to the Oregon Health Plan (OHP)
Standard program, an extension of Medicaid for adults not eligible
for OHP Plus.\footnote{The OHP Plus exclusively covers children, pregnant women and TANF
families.} Lottery registrants were randomly assigned to win conditional on
the number of household members on the waiting list, and the winners
were required to return an application form. Only 60.82\% of the winners
returned the form on time, and some failed subsequent eligibility
screening. This process generated a self-selection problem, yielding
an endogenous treatment variable (enrollment status) and a valid IV
(lottery status). We consider four health outcomes: ``Physical activities,''
``Depression,'' ``Got all needed medical care,'' and ``Got all
needed dental care.'' The first two outcomes measure physical and
mental health, whereas the latter two measure satisfaction with the
medical care and services outside primary care, respectively. 

Our data comprise three parts: (i) baseline characteristics and lottery
status, which were recorded by the researchers at the time of the
lottery, and they do not contain missing values; (ii) treatment variable
from the initial survey (0m) conducted right after the experiment;
and (iii) outcomes from the final round survey (12m) conducted 12
months later. This one-year gap allows the treatment to take effect.
The survey-measured treatment and outcome variables are all partially
missing. 

\begin{figure}
\begin{centering}
\includegraphics[scale=0.8]{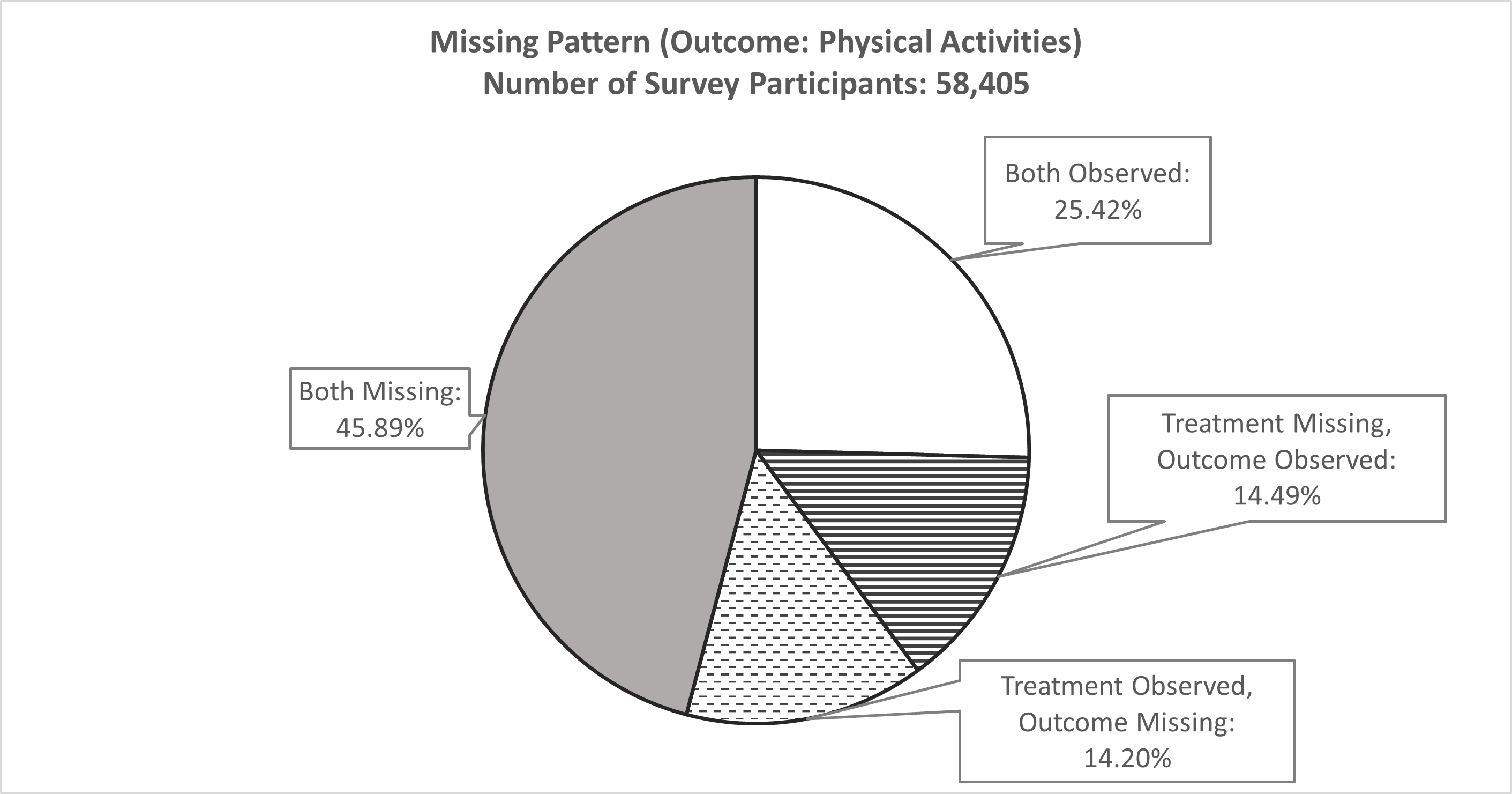}
\par\end{centering}
\caption{Non-monotone Missing Pattern}
\label{fig:=000020non-monotone=000020missing}
\end{figure}

The main source of missingness in this dataset is non-response to
surveys. For both the 0m and 12m surveys, response rates are below
50\%. Among those participants, 16.85\% only participated in the 0m
survey, while 12.32\% responded only to the 12m survey. Another source
of missingness is non-response to certain questions among respondents.
Among the 16,579 participants who returned both the 0m and 12m surveys,
self-reported treatment status remains unconfirmed for 8.84\%. For
the chosen outcome variables, the non-response rates vary from 1.91\%
to 3.60\% among respondents to both surveys. These two sources together
generate a non-monotone missing pattern. Across outcome variables,
approximately 14\%-15\% of participants have the outcome observed
but treatment missing, representing a substantial fraction of the
sample. Figure \ref{fig:=000020non-monotone=000020missing} illustrates
this pattern for one of the outcomes ``Physical Activities,'' demonstrating
a typical non-monotone missing pattern; the other outcomes show similar
proportions.

\section{A Sequential Missing at Random Assumption \protect\label{subsec:A-Sequential-Missing}}

The OHIE example introduced above illustrates a common empirical pattern
of sequentially collected data with multivariate missingness. The
standard strengthened MAR tends to fail in such scenarios, and we
therefore propose a new identification assumption. 

In such settings, it is useful to distinguish the benchmark MAR assumption
from the stronger restrictions that are often imposed with multiple
missing variables. Because the term MAR is used inconsistently across
the literature, we fix terminology first. Throughout this section,
we use the \textit{everywhere MAR} formulation (\citet{seaman2013meant}),
which requires the probability of each possible missingness pattern
to depend only on the variables observed under that pattern. Formally,
let $Y=(X_{1},X_{2},W)$, where $W$ is fully observed and $X_{1}$
and $X_{2}$ are partially missing. Let $R=(R_{1},R_{2})$. For any
missingness pattern $r=(r_{1},r_{2})$, let $o(Y,r)$ denote the subvector
of $Y$ observed under pattern $r$. Everywhere MAR states that
\begin{equation}
\Pr[R=r|Y=y]=\Pr[R=r|Y=y^{*}],\forall r,y,y^{*}\text{ such that }o(y,r)=o(y^{*},r).\label{eq:=000020everywhere=000020MAR-1}
\end{equation}
With two partially missing variables, this implies that there exist
functions $\pi_{00},\pi_{10},\pi_{01}$ and $\pi_{11}$ such that
\begin{align*}
\Pr\left[R_{1}=0,R_{2}=0\mid X_{1},X_{2},W\right]=\pi_{00}(W),\;\;\;\;\; & \Pr\left[R_{1}=1,R_{2}=0\mid X_{1},X_{2},W\right]=\pi_{10}(W,X_{1})\\
\Pr\left[R_{1}=0,R_{2}=1\mid X_{1},X_{2},W\right]=\pi_{01}(W,X_{2}), & \Pr\left[R_{1}=1,R_{2}=1\mid X_{1},X_{2},W\right]=\pi_{11}(W,X_{1},X_{2}).
\end{align*}

However, in such case, researchers usually need to adopt a stronger
restriction under which all missingness mechanisms depend on the same
set of fully observed variables. To see the reason, we further define
the first-stage missing propensity
\[
p_{1}(L)=\Pr\left[R_{1}=1\mid L\right],
\]
 where $L$ denotes a vector of variables, and the conditional and
joint probabilities:
\begin{align*}
p_{r_{2}|r_{1}}(L) & =\Pr\left[R_{2}=r_{2}\mid L,R_{1}=r_{1}\right],\\
p_{r_{1}r_{2}}(L) & =\Pr\left[R_{1}=r_{1},R_{2}=r_{2}\mid L\right],\forall(r_{1},r_{2})\in\{0,1\}^{2}.
\end{align*}
Taking two of the joint probabilities above, we can rewrite them as:
\[
\Pr\left[R_{1}=0,R_{2}=0\mid X_{1},X_{2},W\right]=\left(1-p_{1|0}(X_{1},X_{2},W)\right)\left(1-p_{1}(X_{1},X_{2},W)\right)=\pi_{00}(W),
\]
\[
\Pr\left[R_{1}=1,R_{2}=0\mid X_{1},X_{2},W\right]=\left(1-p_{1|1}(X_{1},X_{2},W)\right)p_{1}(X_{1},X_{2},W)=\pi_{10}(X_{1},W).
\]
These two joint probabilities are built from the same conditional
components but impose different covariate restrictions. For both restrictions
to hold simultaneously, we either need strong structural restrictions
on $p_{1|0}$ and $p_{1}$, or impose the \textit{strengthened MAR}:\footnote{Failure of everywhere MAR with multiple missingness was introduced
by \citet{robins1997non} but they discuss it from the perspective
of estimation challenges. They argue that MAR is difficult to justify
for multivariate missingness, considering a scenario with no fully
observed variable and two partially missing variables, and showing
that, in a logistic model, MAR implicitly collapses to MCAR. It is
also discussed in the following studies (\citet{robins1997non2,little2002bayes,tsiatis2007semiparametric,tchetgen2018discrete}).
Moreover, under such MAR assumption, the associated efficient influence
function generally does not admit closed-form expressions.}
\[
\Pr[R=r|X_{1},X_{2},W]=\tilde{\pi}_{r}(W),\;\;\forall r\in\left\{ (0,0),(1,0),(0,1),(1,1)\right\} .
\]

Our contribution starts from this point. In many applications with
sequential data collection, the strengthened MAR is exactly where
empirical plausibility breaks down. We show later in the OHIE example
that later-stage missingness may depend on earlier partially observed
values. At the same time, allowing unrestricted dependence across
all partially observed variables reintroduces the practical difficulties
that motivated the common-covariate restriction in the first place.
Meanwhile, it is hard to justify how later-realized variables affect
the missingness mechanism in the previous stage. We therefore propose
a middle ground. Our Sequential Missing at Random (SMAR) assumption
preserves tractable propensity-based identification, but relaxes the
symmetric common-covariate restriction in one empirically relevant
direction: later-stage missingness may depend on earlier partially
observed variables on the branch where those variables are observed.

\begin{asSMAR}\label{assn:SMAR}

\begin{align}
 & R_{1}\perp(X_{1},X_{2})\mid W,\label{eq:=000020SMAR-1}\\
 & R_{2}\perp(X_{1},X_{2})\mid W,R_{1}=0,\label{eq:=000020SMAR-2}\\
 & R_{2}\perp X_{2}\mid X_{1},W,R_{1}=1.\label{eq:=000020SMAR-3}
\end{align}

\end{asSMAR}

SMAR is not proposed as a general model for non-monotone missingness.
It is a structured subset of everywhere MAR tailored to identify missingness
mechanisms in some scenarios with a sequential data collection process.
It allows the second-stage missingness mechanism to depend on the
first-stage partially observed variable exactly on the branch where
that variable is observed, while ruling out dependence on variables
that have not yet entered the relevant information set. In this sense,
SMAR is weaker than the strengthened MAR.

The next result shows that SMAR yields a selection-on-observables
representation for the four missingness patterns, denoted by $p_{00},p_{10},p_{01},p_{11}$. 

\begin{proposition}\label{prop:=000020propensity=000020identification}

Under Assumption SMAR,
\begin{align*}
p_{00}(W):=\Pr\left[R_{1}=0,R_{2}=0\mid X_{1},X_{2},W\right] & =\left(1-p_{1|0}(W)\right)\cdot\left(1-p_{1}(W)\right),\\
p_{10}(X_{1},W):=\Pr\left[R_{1}=1,R_{2}=0\mid X_{1},X_{2},W\right] & =\left(1-p_{1|1}(X_{1},W)\right)\cdot p_{1}(W),\\
p_{01}(W):=\Pr\left[R_{1}=0,R_{2}=1\mid X_{1},X_{2},W\right] & =p_{1|0}(W)\cdot\left(1-p_{1}(W)\right),\\
p_{11}(X_{1},W):=\Pr\left[R_{1}=1,R_{2}=1\mid X_{1},X_{2},W\right] & =p_{1|1}(X_{1},W)\cdot p_{1}(W).
\end{align*}

\end{proposition}

Proposition \ref{prop:=000020propensity=000020identification} is
useful because it converts the non-monotone missingness into three
propensity functions that are identifiable from observed data. The
first-stage propensity $p_{1}(W)$ is identifiable from the full sample.
$p_{1|0}(W)$ and $p_{1|1}(X_{1},W)$ are correspondingly identified
in the subsamples $R_{1}=0$ and $R_{1}=1$, respectively. The conditioning
set in each propensity contains only variables observed in the relevant
comparison: $W$ for the first-stage propensity and the $R_{1}=0$
branch, and $(X_{1},W)$ for the $R_{1}=1$ branch. Thus, SMAR can
be viewed as the least restrictive stagewise propensity specification
compatible with the sequential order of data collection, and this
representation makes SMAR directly compatible with standard IPW and
AIPW implementation.

To motivate this assumption, it is useful to think in terms of information
sets. $R_{1}$ can measure whether the first-stage value $X_{1}$
has entered the information set of the respondent, the interviewer,
or the data-collection process by the time $R_{2}$ is determined.
If later response may depend on previously revealed first-stage values,
but not on values that have not been revealed along that branch, then
an asymmetric restriction becomes natural. We provide two examples.
The first one illustrates respondent-side information acquisition.
The second one highlights how asymmetric missingness can be caused
by researcher-side asymmetric information, naturally arising from
experimental design.

\begin{example}[\textbf{Respondent-side asymmetric information}]

Consider a panel where a parent is asked in two successive semesters
to report their children's grades ($X_{t}$). Suppose parents are
``honest'' and answer whenever they know the grade. Then $R_{1}=1$
means that the first-semester grade has entered the parent's information
set. For these parents, the realized value of $X_{1}$ may affect
subsequent monitoring effort such that a poor grade may induce closer
attention next semester, whereas a strong grade may induce complacency.
As a result, second-stage response rate depends on $X_{1}$ within
the $R_{1}=1$ group. By contrast, when $R_{1}=0$, the parent did
not know the first-semester grade, and the latent value of $X_{1}$
cannot itself affect later period reporting because it never entered
the parent's information set. For this branch, second-stage missingness
is independent of the unknown previous grade. 

\end{example}

\begin{example}[\textbf{Researcher-side asymmetric information}]

Consider a field experiment in which individuals are randomly assigned
cash vouchers for flu vaccination.\footnote{This is a real experiment registered on AEA RCT Registry with the
ID AEARCTR-0016944.} Let $X_{1}$ denote the take-up decision and $X_{2}$ be the outcomes
collected in subsequent surveys. Participants self-report their take-up
decision,\footnote{To guarantee both control and treated group are willing to report
their decision, we set extra survey bonus.} and vaccinated participants can redeem the voucher via a reimbursement
system by uploading supportive documents. In this setting $R_{1}=1$
indicates that take-up status has entered the research team's information
set. Among these units, the research team has closer contact with
vaccinated groups to assist subsequent redemption. Many in the treated
group have joined a Q\&A group chat where survey reminders are also
published. Such additional contact can affect later survey participation.
Therefore, later missingness may depend on the realized value of $X_{1}$.
By contrast, the research team does not observe take-up status for
those with $R_{1}=0$ and cannot condition subsequent contact on the
unobserved treatment status. In this branch, latent $X_{1}$ should
not affect $R_{2}$ through researcher-side follow-up. 

\end{example}

Importantly, we do not propose SMAR as a general model for non-monotone
missingness. Rather, SMAR is intended for a subset of applications
in which the order of data collection is substantively meaningful
and provides a credible basis for asymmetric dependence in the missingness
mechanism. We need to determine which identification assumption to
use based on the features of the data. If the missing variables are
collected simultaneously and there is no obvious ordering of variables,
one should instead work directly with strengthened MAR or another
application-specific identifying assumption.

Reassuringly, the approach we will propose later works as well under
the strengthened MAR, as discussed below. For now, we show some evidence
that SMAR is more plausible than strengthened MAR with the OHIE example,
as further justification. 

\subsection{Evidence on the Plausibility of SMAR in the OHIE Data\protect\label{subsec:Running-Example-(continue):}}

We now return to the OHIE survey data to assess the plausibility of
the SMAR assumption. Like any MAR-type assumption, SMAR is not testable.
Our goal here is therefore more limited. We use simple regressions
to show that: (i) strengthened MAR does not hold in the OHIE data;
(ii) the observable implications of SMAR hold. We borrow a fully observed
proxy variable for $X_{1}$ from the administrative data to provide
evidence on asymmetric dependence between $R_{2}$ and $X_{1}$ depending
on $X_{1}$'s observability, which is difficult to assess using only
observed data. All regressions control for the same set of covariates
$W$ used in the main analysis.

We begin by documenting evidence against the strengthened MAR assumption.
In particular, Table \ref{tab: evidence1} shows that when the first-stage
treatment $X_{1}$ is observed, there is a significant correlation
between it and the second-stage missing mechanisms $R_{2}$ for all
four outcome variables. This finding contradicts the strengthened
MAR assumption, under which $R_{2}$ should be independent of $X_{1}$
conditional on $W$. We could certainly argue that the independence
condition could still hold statistically at the unconditional level,
though conditional on $R_{1}=1$ it does not hold. However, such an
interpretation is hard to justify and requires strong assumptions
on the distributional structure. Therefore, we see it as evidence
that strengthened MAR fails in this dataset.

\begin{table} 
\caption{Regression of $R_2$ on $X_1$ when $X_1$ is observed} 
\label{tab: evidence1}
\def\sym#1{\ifmmode^{#1}\else\(^{#1}\)\fi}
\resizebox{\textwidth}{!}{
\begin{tabular}{l*{5}{c}} 
\hline\hline                     
&\multicolumn{1}{c}{Physical}&\multicolumn{1}{c}{}&\multicolumn{1}{c}{Got All Needed}&\multicolumn{1}{c}{Got All Needed}\\ 
&\multicolumn{1}{c}{Activities}&\multicolumn{1}{c}{Depression}&\multicolumn{1}{c}{Medical Care}&\multicolumn{1}{c}{Dental Care}\\ 
\hline 
Currently have OHP insurance                          &   -0.0373\sym{***}  &   -0.0371\sym{***}  &   -0.0388\sym{***} &   -0.0391\sym{***}\\         
                                                      &   (0.00855)         &   (0.00858)         &    (0.00864)       &   (0.00860)\\ 
[1em] Control variables                               &   Yes  &   Yes  &   Yes &   Yes\\                                    [1em] 
\hline Observations                                   &     23140           &     23140           &      23140         &    23140\\ 
\hline\hline 
\multicolumn{5}{l}{\footnotesize Control variables include: Selected in the lottery; Number of people in household; Female; Age; Zip code in a }\\
\multicolumn{5}{l}{\footnotesize metropolitan statistical area; Individual requested English-language materials. Standard errors in parentheses.}\\ 
\multicolumn{5}{l}{\footnotesize \sym{*} \(p<0.05\), \sym{**} \(p<0.01\), \sym{***} \(p<0.001\).}\\
\end{tabular} 
}
\end{table}

Such correlation with a partially observed variable does not appear
for $R_{1}$. We run an analogous regression of $R_{1}$ on $X_{2}$
when $X_{2}$ is observed, and find negligible and statistically insignificant
correlations. This is caused by the time gap between $R_{1}$ and
$X_{2}$. Since $R_{1}$ is realized before $X_{2}$, it is reasonable
to assume later-stage variables, regardless of their observability,
could not affect the previous-stage missingness mechanism. The results
are presented in Table \ref{tab: evidence2}.

\begin{table} 
\caption{Regression of $R_1$ on $X_2$ when $X_2$ is observed}
\label{tab: evidence2}
\def\sym#1{\ifmmode^{#1}\else\(^{#1}\)\fi} 
\resizebox{\textwidth}{!}{
\begin{tabular}{l*{5}{c}} 
\hline\hline                     
&\multicolumn{1}{c}{Physical}&\multicolumn{1}{c}{}&\multicolumn{1}{c}{Got All Needed}&\multicolumn{1}{c}{Got All Needed}\\ 
&\multicolumn{1}{c}{Activities}&\multicolumn{1}{c}{Depression}&\multicolumn{1}{c}{Medical Care}&\multicolumn{1}{c}{Dental Care}\\ 
\hline 
Physical activities                              &   -0.000825         &                     &                   &            \\        
                                                       &   (0.00429)         &                     &                   &            \\ 
[1em] Depression                                       &                     &   -0.00557          &                   &            \\       
                                                       &                     &   (0.00323)         &                   &             \\ 
[1em] Got all needed medical care                      &                     &                     &    0.0114         &             \\       
                                                       &                     &                     &    (0.00657)      &              \\
[1em] Got all needed dental care                       &                     &                     &                   &    0.00578   \\                                                                     &                     &                     &                  &    (0.00645)\\
Control variables                                &   Yes          &   Yes          &   Yes        &   Yes      \\ 
[1em]\hline 
Observations                                           &    23308            &   23181             &    22940          &    23172\\ 
\hline\hline 
\multicolumn{5}{l}{\footnotesize Control variables include: Selected in the lottery; Number of people in household; Female; Age; Zip code in a }\\
\multicolumn{5}{l}{\footnotesize metropolitan statistical area; Individual requested English-language materials. Standard errors in parentheses.}\\ 
\multicolumn{5}{l}{\footnotesize \sym{*} \(p<0.05\), \sym{**} \(p<0.01\), \sym{***} \(p<0.001\).}\\
\end{tabular}
}
\end{table} 

The results in the above two tables provide some evidence against
strengthened MAR, and support some observable implications of SMAR.
But the most crucial implication needs to be verified is the asymmetric
relationship between $X_{1}$ and $R_{2}$ depending on whether $X_{1}$
is observed at the first stage. When $R_{1}=1$, $R_{2}$ may depend
on $X_{1}$; however, this dependence should no longer be present
when $R_{1}=0$. This is the key implication we examine next.

In Table \ref{tab: evidence1}, we have shown that $R_{2}$ correlates
with $X_{1}$ when $R_{1}=1$. However, since the survey treatment
variable $X_{1}$ is missing when $R_{1}=0$, we cannot examine this
implication directly using the survey measure itself. We therefore
borrow administratively-recorded Medicaid enrollment status from \citet{finkelstein2012oregon}
as a proxy for survey-collected enrollment status. This proxy is observed
for the full sample, which allows us to examine the association between
treatment and $R_{2}$ separately by the observability of the survey
treatment variable. We first consider the subsample with observed
treatment $R_{1}=1$ to verify the pattern we found in Table \ref{tab: evidence1},
and to validate the use of the proxy variable. In Table \ref{tab: evidence3},
we show that the administrative proxy has a relationship with $R_{2}$
very similar to that of the survey treatment variable: among individuals
whose treatment is observed, treatment status is significantly correlated
with the second-stage missingness.

We then turn to the most important step and use the subsample with
missing survey treatment ($R_{1}=0$). In this group, we regress $R_{2}$
on the proxy for $X_{1}$. Table \ref{tab: evidence4} shows no meaningful
or statistically significant correlation between the administrative
proxy and $R_{2}$. This contrast across the two subsamples is the
main empirical support for SMAR in the OHIE data. The treatment-missingness
link is present when treatment is observed at the first stage, and
absent when treatment is not observed. 

\begin{table} 
\caption{Regression of $R_2$ on proxy of $X_1$ when $X_1$ is observed} 
\label{tab: evidence3}
\def\sym#1{\ifmmode^{#1}\else\(^{#1}\)\fi} 
\resizebox{\textwidth}{!}{
\begin{tabular}{l*{5}{c}} 
\hline\hline                     
&\multicolumn{1}{c}{Physical}&\multicolumn{1}{c}{}&\multicolumn{1}{c}{Got All Needed}&\multicolumn{1}{c}{Got All Needed}\\ 
&\multicolumn{1}{c}{Activities}&\multicolumn{1}{c}{Depression}&\multicolumn{1}{c}{Medical Care}&\multicolumn{1}{c}{Dental Care}\\
\hline 
Currently have OHP insurance                            &     -0.0233\sym{**} &    -0.0241\sym{***} &    -0.0253\sym{***} &    -0.0272\sym{***}\\                                                            
(proxy) &     (0.00725)       &    (0.00727)        &    (0.00732)        &    (0.00729) \\ 
[1em] Control variables                               &   Yes  &   Yes  &   Yes &   Yes\\
[1em]\hline Observations                                     &      23140          &     23140           &     23140           &     23140\\ 
\hline\hline 
\multicolumn{5}{l}{\footnotesize Control variables include: Selected in the lottery; Number of people in household; Female; Age; Zip code in a }\\
\multicolumn{5}{l}{\footnotesize metropolitan statistical area; Individual requested English-language materials. Standard errors in parentheses.}\\ 
\multicolumn{5}{l}{\footnotesize \sym{*} \(p<0.05\), \sym{**} \(p<0.01\), \sym{***} \(p<0.001\).}\\
\end{tabular} 
}
\end{table}

\begin{table} 
\caption{Regression of $R_2$ on proxy of $X_1$ when $X_1$ is not observed} 
\label{tab: evidence4}
\def\sym#1{\ifmmode^{#1}\else\(^{#1}\)\fi} 
\resizebox{\textwidth}{!}{
\begin{tabular}{l*{5}{c}} 
\hline\hline                     
&\multicolumn{1}{c}{Physical}&\multicolumn{1}{c}{}&\multicolumn{1}{c}{Got All Needed}&\multicolumn{1}{c}{Got All Needed}\\ 
&\multicolumn{1}{c}{Activities}&\multicolumn{1}{c}{Depression}&\multicolumn{1}{c}{Medical Care}&\multicolumn{1}{c}{Dental Care}\\
\hline 
Currently have OHP insurance                            &      0.00528        &     0.00491         &    0.00492          &    0.00499\\
(proxy)                                                 &     (0.00543)       &    (0.00542)        &    (0.00541)        &    (0.00543) \\ 
[1em] Control variables                               &   Yes  &   Yes  &   Yes &   Yes\\
[1em]\hline 
Observations                                     &     35265           &     35265           &     35265           &     35265\\ 
\hline\hline 
\multicolumn{5}{l}{\footnotesize Control variables include: Selected in the lottery; Number of people in household; Female; Age; Zip code in a }\\
\multicolumn{5}{l}{\footnotesize metropolitan statistical area; Individual requested English-language materials. Standard errors in parentheses.}\\ 
\multicolumn{5}{l}{\footnotesize \sym{*} \(p<0.05\), \sym{**} \(p<0.01\), \sym{***} \(p<0.001\).}\\
\end{tabular} 
}
\end{table}

To summarize, the empirical evidence above shows that: (i) $R_{2}$
correlates with $X_{1}$ when $R_{1}=1$ (failure of strengthened
MAR; support of SMAR (3.4)); (ii) $R_{2}$ does not correlate with
$X_{1}$ when $R_{1}=0$ (support of SMAR (3.3)); and (iii) $R_{1}$
does not correlate with partially observed $X_{2}$ (support of SMAR
(3.2)).\footnote{On the outcome side, our argument does not rely on an additional proxy
exercise because it is difficult to obtain proxies for subjective
health measures. Therefore, it is difficult to further show independence
between $X_{2}$ and $R_{1}$ when $X_{2}$ is unobserved. However,
we think it is easy to justify this restriction because a later-realized
variable usually cannot influence the early-stage missingness mechanism.
Prediction models such as Roy model are exceptions, but in such models,
unrealized outcome affects treatment instead of missingness of treatment.
Moreover, we have seen evidence that $R_{1}$ and $X_{2}$ are uncorrelated
conditional on $W$ in Table \ref{tab: evidence2}. The relevant restriction
is the standard one used under MAR in a sequential setting: once W
is controlled for, the later outcomes should not display a systematic
association with the earlier missingness mechanism. It shows that
the data do not suggest a comparable outcome-side pattern. The distinctive
empirical content of SMAR in this application therefore lies in the
asymmetric relationship between $X_{1}$ and $R_{2}$, rather than
in an additional assumption about $X_{2}$.} Taken together, these results do not prove SMAR, as it is untestable,
and they do not address all components of standard MAR-type restrictions
such as the independence between $X_{1}$ and $R_{1}$, as well as
$X_{2}$ and $R_{2}$. What they show is the OHIE dataset follows
the asymmetric dependence pattern that SMAR is designed to capture:
we find that second-stage missingness is related to treatment when
treatment is observed at the first stage, and that this relationship
disappears when treatment is not observed. This asymmetric pattern
is difficult to reconcile with a simple MAR benchmark and provides
direct empirical support for the plausibility of SMAR in this application.
The mechanisms behind these patterns are consistent with the discussion
in the program-status example above. The missing treatment status
may reflect delayed or unconfirmed enrollment information. Using an
external administrative measure for comparison, we find that some
individuals acknowledged successful enrollment only after substantial
delay, up to 277 days.

\section{The AIPW-GMM Estimator and Statistical Properties\protect\label{sec:The-AIPW-GMM-Estimator}}

IPW is a standard way to recover population moments in the presence
of missing variables (\citet{rosenbaum1983central,wooldridge2007inverse,seaman2013review}).
It eliminates bias by consistently reweighting complete observations
by the inverse observation probability, under a MAR-type assumption,
but generally does not recover the efficiency loss from dropping observations
with incomplete variables. Augmenting the IPW score with suitable
augmentation terms yields an augmented IPW (AIPW) moment that is well
known to be semiparametrically efficient and double robust under suitable
conditions (\citet{robins1994estimation}; \citet{robins1997non2};
\citet{carpenter2006comparison}; \citet{tsiatis2007semiparametric};
\citet{chen2008semiparametric}; \citet{glynn2010introduction}). 

The main difficulty in our setting is that the missingness pattern
is non-monotone and the relevant information set differs across missingness
states. As a result, the key mechanism that yields a closed-form efficient
influence function breaks down (\citet{tsiatis2007semiparametric,chaudhuri2016gmm}),
and it is difficult to obtain a suitable augmentation term for the
AIPW moment. Our construction addresses this by using asymmetric augmentation
terms tailored to the specific information available under each missingness
pattern. This asymmetry is the key feature that allows the estimator
to remain robust under SMAR while still admitting a closed-form orthogonal
moment. 

\subsection{AIPW Moment Condition \protect\label{subsec:AIPW-Moment-Condition}}

For notational simplicity, write $p_{1}\equiv p_{1}(W)$, $p_{11}\equiv p_{11}(X_{1},W)$,
$p_{01}\equiv p_{01}(W)$, and denote the observed variables as $O\equiv\left(R_{1},R_{2},R_{1}X_{1},R_{2}X_{2},W\right)$.
Define the nuisance regression functions
\begin{align*}
\mu(W;\beta)=\E\left[g(X,W;\beta)\mid W\right],\;\; & \mu_{1}(W;\beta)=\E\left[g_{1}(X_{1},W;\beta)\mid W\right],\\
\mu_{20}(W;\beta)=\E\left[g_{2}(X_{2},W;\beta)\mid W\right],\;\; & \mu_{21}(X_{1},W;\beta)=\E\left[g_{2}(X_{2},W;\beta)\mid X_{1},W\right],
\end{align*}
and collect them with the propensity components in $\eta(\beta)=\left(p_{1},p_{11},p_{01},\mu,\mu_{1},\mu_{20},\mu_{21}\right)$.

We define the AIPW moment by
\begin{equation}
g_{aipw}(O;\beta,\eta)=\frac{R_{1}R_{2}}{p_{11}}g(X,W;\beta)+\phi(O;\beta,\eta),\label{eq:=000020aipw=000020moment=000020function}
\end{equation}
where the augmentation term is 
\begin{align}
\phi(O;\beta,\eta) & =\left(1-\frac{R_{1}R_{2}}{p_{11}}\right)\mu(W;\beta)\nonumber \\
 & \;\;+\left(\frac{R_{1}}{p_{1}}-\frac{R_{1}R_{2}}{p_{11}}\right)\left(g_{1}(X_{1},W;\beta)-\mu_{1}(W,\beta)\right)\nonumber \\
 & \;\;+p_{1}\cdot\left(\frac{R_{1}}{p_{1}}-\frac{R_{1}R_{2}}{p_{11}}\right)\left(\mu_{21}(X_{1},W;\beta)-\mu_{20}(W;\beta)\right)\nonumber \\
 & \;\;+(1-p_{1})\cdot\left(\frac{(1-R_{1})R_{2}}{p_{01}}-\frac{R_{1}R_{2}}{p_{11}}\right)\left(g_{2}(X_{2},W;\beta)-\mu_{20}(W;\beta)\right).\label{eq:=000020augmentation}
\end{align}
Recall that we impose additive separability on the moment function
in equation \ref{eq:=000020model}, so that $X_{1}$ and $X_{2}$
enter the moment condition through $g_{1}$ and $g_{2}$ separately.
The augmentation can therefore exploit different information sets
for the $X_{1}$- and $X_{2}$- components separately. This is a crucial
feature for deriving closed-form efficiency bound under asymmetric
missingness mechanisms. The last two terms capture distinct information
sets for imputing $X_{2}$ depending on whether $X_{1}$ is observed.
In contrast, $X_{2}$ cannot be used to impute $X_{1}$ in reverse,
because the observability of $X_{2}$ may be correlated with $X_{1}$. 

If we reorganize the augmentation term above and further define
\[
\mu(X_{1},W;\beta)=\E\left[g(X,W;\beta)\mid X_{1},W\right],
\]
we can also write it as
\begin{align*}
\phi(O;\beta,\eta) & =\left(1-\frac{R_{1}R_{2}}{p_{11}}\right)\mu(W;\beta)\\
 & \;\;+\left(\frac{R_{1}}{p_{1}}-\frac{R_{1}R_{2}}{p_{11}}\right)\left(\mu(X_{1},W;\beta)-\mu(W;\beta)\right)\\
 & \;\;+\left(1-p_{1}\right)\left\{ \left(\frac{\left(1-R_{1}\right)R_{2}}{p_{01}}-\frac{R_{1}R_{2}}{p_{11}}\right)\left(g_{2}(X_{2},W;\beta)-\mu_{20}(W;\beta)\right)\right.\\
 & \;\;\left.-\left(\frac{R_{1}}{p_{1}}-\frac{R_{1}R_{2}}{p_{11}}\right)\left(\mu_{21}(X_{1},W;\beta)-\mu_{20}(W;\beta)\right)\right\} .
\end{align*}
In this representation, the first two components coincide with the
augmentation terms for monotone missingness proposed by \citet{chaudhuri2020efficiency}.
The last two components additionally incorporate observations with
$R_{1}=0$ and $R_{2}=1$, i.e., the strictly non-monotone missing
part.

The nuisance functions appearing in $g_{aipw}$ are of two types:
missingness propensities $p_{1}$, $p_{01}$, $p_{11}$ and the conditional
expectation functions $\mu$, $\mu_{1}$, $\mu_{20}$, $\mu_{21}$.
Identification of the propensity components was established in the
previous section. The imputation components are identified as follows.

\begin{proposition}\label{prop:=000020augementation=000020identification}

When Assumption SMAR holds, 
\begin{align*}
 & \mu_{1}(W;\beta)=\E\left[g_{1}(X_{1},W;\beta)\mid W,R_{1}=1\right],\\
 & \mu_{20}(W;\beta)=\E\left[g_{2}(X_{2},W;\beta)\mid W,R_{1}=0,R_{2}=1\right],\\
 & \mu_{21}(X_{1},W;\beta)=\E\left[g_{2}(X_{2},W;\beta)\mid X_{1},W,R_{1}=1,R_{2}=1\right],\\
 & \mu(W;\beta)=\mu_{1}(W;\beta)+\E\left[\mu_{21}(X_{1},W;\beta)|W\right].
\end{align*}

\end{proposition}

We next introduce a standard overlap condition to ensure the moment
function above is well-defined.

\begin{asOVERLAP}\label{assn:=000020overlap}

There exists $c>0$ such that $p_{1},p_{01}\geq c$ almost surely
in $W$, $p_{11}\geq c$ almost surely in $(W,X_{1})$. 

\end{asOVERLAP}

\begin{theorem}\label{thm:=000020identification}

Suppose Assumption SMAR and Overlap hold, and let $\eta_{0}(\beta)$
denote the collection of true propensity and imputation functions.
Then
\[
\E\left[g_{aipw}(O;\beta,\eta_{0}(\beta))\right]=0\text{ if and only if }\beta=\beta^{0}.
\]

\end{theorem}

\subsection{Double Robustness\protect\label{subsec:Double-Robustness}}

The AIPW moment is well known for its double robustness property.
In general, misspecification of first-stage nuisance functions can
induce bias through incorrect functional forms, omitted variables,
or measurement error, to name a few. In our framework, misspecification
of either the missingness propensities or the imputation functions
can be allowed as long as the other part is correctly specified. 

We allow a broad class of misspecifications but impose one restriction
on propensity models under SMAR: they must remain functions of the
appropriate information sets. Specifically, $p_{1}$ and $p_{01}$
can only be misspecified as functions of $W$ or any subvector of
$W$, and $p_{11}$ is specified as a function of variables contained
in $(X_{1},W)$. This restriction allows us to separate the imputation
terms from the propensities in expectations, so that correctness of
the imputation regressions alone suffices for consistency, even under
misspecification. This still permits the standard forms of misspecification
that are most relevant in practice, such as omitted variables within
the admissible information set, incorrect functional form, or an incorrect
link function. One familiar example is when a logit model is used
although the true response probabilities follow a probit model, or
conversely. 

In contrast, we place no analogous restriction on misspecified imputation
models for the double-robustness claim. This is because, even when
conditioning on $X_{2}$, SMAR implies $\E\left[R_{1}R_{2}|X_{1},X_{2},W\right]=p_{11}(X_{1},W)$
, which preserves the required centering. Full derivations are provided
in the Appendix. We formally define the double robustness as follows.

\begin{theorem}\label{thm:=000020robustness=000020SMAR}

Let $\overline{\eta}(\beta)=(\overline{p}_{1},\overline{p}_{11},\overline{p}_{01},\overline{\mu},\overline{\mu}_{1},\overline{\mu}_{20},\overline{\mu}_{21})$
denote any collection of measurable functions such that $\overline{p}_{1}$
and $\overline{p}_{01}$ depend only on $W$, $\overline{p}_{11}$
is a function of $(X_{1},W)$, all three propensity components are
positive and bounded away from zero, and the imputation components
are square-integrable. Suppose Assumption SMAR and Overlap hold. Then
the identification statement in Theorem \ref{thm:=000020identification}
holds for $\overline{\eta}$ in either of the following cases:
\begin{itemize}
\item [(a)](Propensities correct) the missingness propensities $(\overline{p}_{1},\overline{p}_{11},\overline{p}_{01})$
are correctly specified, while $(\overline{\mu},\overline{\mu}_{1},\overline{\mu}_{20},\overline{\mu}_{21})$
may be misspecified. 
\item [(b)](Imputations correct) $(\overline{\mu},\overline{\mu}_{1},\overline{\mu}_{20},\overline{\mu}_{21})$
are correctly specified, while $(\overline{p}_{1},\overline{p}_{11},\overline{p}_{01})$
may be misspecified provided they maintain the information sets above
(i.e., $\overline{p}_{1}$ and $\overline{p}_{01}$ are specified
as functions of variables contained in $W$; $\overline{p}_{11}$
as a function of variables contained in $(W,X_{1})$ ) and remain
positive and bounded away from zero.
\end{itemize}
\end{theorem}

Theorem \ref{thm:=000020robustness=000020SMAR} is a global population
result. It states that the AIPW moment remains correctly centered
at the true value even when some nuisance parameters are misspecified.
A related concept is Neyman orthogonality. However, it plays a different
role. It is a local derivative statement used for asymptotic theory.
Combined with some other regularity conditions discussed in the later
section, Neyman orthogonality guarantees that the first-stage estimation
does not enter the first-order asymptotic distribution of the second-step
estimator and ensures efficiency of the second-stage estimation. 

To state orthogonality precisely, let $p^{0}=(p_{1}^{0},p_{01}^{0},p_{11}^{0})$
and $m^{0}=(\mu^{0},\mu_{1}^{0},\mu_{20}^{0},\mu_{21}^{0})$ denote
the true values of the nuisance terms, and $\eta^{0}=(p^{0},m^{0})$;
let $H_{p}$ denote the class of square-integrable perturbations $h_{p}=(h_{p_{1}},h_{p_{01}},h_{p_{11}})$
such that $p^{0}+th_{p}$ remains admissible for all sufficiently
small $t$; let $H_{m}$ denote the class of square-integrable perturbations
$h_{m}=(h_{\mu},h_{\mu_{1}},h_{\mu_{20}},h_{\mu_{21}})$ respecting
the conditioning structures of $m^{0}$. We claim Neyman orthogonality
below.

\begin{corollary}\label{lem:=000020neyman-ortho}

Suppose Assumption SMAR and the Overlap condition hold, and assume
that for every $h_{p}\in H_{p}$ and $h_{m}\in H_{m}$ the maps
\begin{align*}
t\mapsto\E\left[g_{aipw}\left(O;\beta^{0},(p^{0}+th_{p},m^{0})\right)\right], & t\mapsto\E\left[g_{aipw}\left(O;\beta^{0},(p^{0},m^{0}+th_{m})\right)\right]
\end{align*}
are differentiable at $t=0$. Then
\begin{align*}
\frac{d}{dt}\E\left[g_{aipw}\left(O;\beta^{0},(p^{0}+th_{p},m^{0})\right)\right]|_{t=0}=\frac{d}{dt}\E\left[g_{aipw}\left(O;\beta^{0},(p^{0},m^{0}+th_{m})\right)\right]|_{t=0}=0
\end{align*}
Equivalently, the AIPW moment is Neyman-orthogonal with respect to
the nuisance functions at the truth.

\end{corollary}

The proof follows directly from Theorem \ref{thm:=000020robustness=000020SMAR}.
Indeed, Theorem \ref{thm:=000020robustness=000020SMAR} implies the
blockwise identities $\E\left[g_{aipw}(O;\beta^{0},(p,m^{0}))\right]=0$
for every admissible $p$ and $\E\left[g_{aipw}(O;\beta^{0},(p^{0},m)\right]=0$
for every admissible $m$. Differentiating these identities at the
true values yields the stated directional derivatives above. Therefore,
in the present setting, orthogonality is the local implication of
the double-robust identities. 

A byproduct of this approach is that the above results also hold under
strengthened MAR assumption, if we correctly define the missingness
propensities, i.e., all propensities are specified as functions of
$W$. A direct result is that the desirable statistical properties
holding under SMAR also hold under strengthened MAR. A more detailed
discussion is included in the Appendix.

\subsection{Estimation and Asymptotic Properties\protect\label{subsec:Efficient-Estimator}}

Following the moment condition above, we estimate $\beta^{0}$ by
plug-in GMM. Before doing so, we first introduce the GMM regularity
conditions in our setting.

\begin{asM}\label{assn:=000020GMM}
\begin{itemize}
\item [(1)]$(X_{1,i},X_{2,i},W_{i},R_{1,i},R_{2,i},R_{1,i}X_{1,i},R_{2,i}X_{2,i})$
are i.i.d. across $i$;
\item [(2)]$g(X,W;\beta)$ is continuously differentiable with respect
to $\beta\in int(\mathcal{B})$, $\E\left[||g(X,W;\beta)||^{4}\right]<\infty$,
$\sup_{\beta\in\mathcal{B}}\E\left[||\frac{\partial}{\partial\beta'}g(X,W;\beta)||^{4}\right]<\infty$,
and there exist measurable envelope functions, $\overline{M}_{1}(X_{1},W)$,
$\overline{M}_{2}(X_{2},W)$ and $\overline{M}_{\partial g}(X,W)$
with $E\left[\overline{M}_{1}^{4}\right]<\infty,E\left[\overline{M}_{2}^{4}\right]<\infty,E\left[\overline{M}_{\partial g}^{4}\right]<\infty$,
such that $||g_{1}(X_{1},W;\beta)||\leq\overline{M}_{1}(X_{1},W)$,
$||g_{2}(X_{2},W;\beta)||\leq\overline{M}_{2}(X_{2},W)$ and $||\frac{\partial}{\partial\beta'}g(X,W;\beta)||\leq\overline{M}_{\partial g}(X,W)$;
\item [(3)]Define $G(\beta)\equiv\frac{\partial}{\partial\beta^{'}}\E\left[g(X,W;\beta)\right]$.
$G(\beta)$ has full rank at $\beta=\beta^{0}$;
\item [(4)]$\operatorname{Var}\left(g(X,W;\beta)\right)$ and $\operatorname{Var}\left(\phi(O;\beta,\eta(\beta))\right)$
are bounded and positive definite for $\beta\in\mathcal{B}$ and are
positive definite at $\beta=\beta^{0}$.
\end{itemize}
\end{asM}

The GMM regularity conditions are imposed on the original structural
moment $g(X,W;\beta)$. Beyond the standard regularity conditions
required for GMM, we add an extra envelope condition in M(2) to ensure
that the relevant conditional expectations are also smooth, so analogous
regularity conditions carry over to $g_{aipw}$. We implement an efficient
GMM estimator in three steps.
\begin{itemize}
\item [\textbf{Step 1}] 

Construct appropriate estimators for missingness propensities $\hat{p}\equiv(\hat{p}_{1},\hat{p}_{01},\hat{p}_{11})$
and the imputation functions $\hat{m}(\beta)\equiv(\hat{\mu},\hat{\mu}_{1},\hat{\mu}_{20},\hat{\mu}_{21})$.
For any $\beta$, define the sample AIPW moment by
\[
\hat{g}_{aipw}(\beta;\hat{p},\hat{m}(\beta))=\frac{1}{n}\sum_{i=1}^{n}g_{aipw}\left(O_{i};\beta,\hat{p},\hat{m}(\beta)\right),
\]
where $g_{aipw}\left(O_{i};\beta,\hat{p},\hat{m}(\beta)\right)$ is
the population AIPW moment with the nuisance functions replaced by
their first-step estimators.
\item [\textbf{Step 2}] 

Let $\hat{\Lambda^{0}}$ be any symmetric positive definite weighting
matrix such that $\hat{\Lambda^{0}}\rightarrow_{p}\Lambda^{0}$, where
$\Lambda^{0}$ is also symmetric positive definite. We first obtain
a preliminary $\hat{\beta}^{(1)}$ from
\[
\hat{\beta}^{(1)}=argmin_{\beta}\hat{g}_{aipw}(\beta;\hat{p},\hat{m}(\beta))^{'}\hat{\Lambda^{0}}\hat{g}_{aipw}(\beta;\hat{p},\hat{m}(\beta)).
\]

Here we use the superscript $(1)$ to indicate that this is the initial
estimator of $\beta^{0}$ that we use for variance calculation. When
the model is just identified, this step is equivalent to solving $\hat{g}_{aipw}=0$,
but the GMM formulation is retained here to cover the overidentification
case.
\item [\textbf{Step 3}] 

Using the preliminary estimator $\hat{\beta}^{(1)}$, we can construct
a consistent estimator of $V=\Var\left(g_{aipw}(O;\beta^{0})\right)$
by the sample analogue
\[
\hat{V}^{(1)}=\frac{1}{n}\sum_{i=1}^{n}g_{aipw}\left(O_{i};\hat{\beta}^{(1)},\hat{p},\hat{m}(\hat{\beta}^{(1)})\right)g_{aipw}\left(O_{i};\hat{\beta}^{(1)},\hat{p},\hat{m}(\hat{\beta}^{(1)})\right)^{'}.
\]
The optimal weighting matrix is set $\hat{\Lambda^{*}}=\left(\hat{V}^{(1)}\right)^{-1}.$
The final efficient estimator is
\[
\hat{\beta}=argmin_{\beta}\hat{g}_{aipw}(\beta;\hat{p},\hat{m}(\beta))^{'}\hat{\Lambda^{*}}\hat{g}_{aipw}(\beta;\hat{p},\hat{m}(\beta)).
\]

\end{itemize}
For inference, define the final score covariance estimator
\[
\hat{V}=\frac{1}{n}\sum_{i=1}^{n}g_{aipw}\left(O_{i};\hat{\beta},\hat{p},\hat{m}(\hat{\beta})\right)g_{aipw}\left(O_{i};\hat{\beta},\hat{p},\hat{m}(\hat{\beta})\right)^{'}
\]
and the sample Jacobian
\[
\hat{G}=\frac{\partial}{\partial\beta^{'}}\hat{g}_{aipw}\left(\beta;\hat{p},\hat{m}(\beta)\right)\vert_{\beta=\hat{\beta}}.
\]
The asymptotic variance estimator of $\hat{\beta}$ is $\hat{\Omega^{*}}=(\hat{G}'\hat{V}^{-1}\hat{G})^{-1}.$

These steps complete the estimation procedure. The first step estimates
nuisance functions; the second and third steps are no different from
standard efficient GMM estimation, but we make them explicit since
the primary appeal of the proposed method is its practical applicability.
For implementation, a closed-form $V$ is not needed. It suffices
to define it as the sample covariance matrix of $g_{aipw}\left(O_{i};\hat{\beta},\hat{p},\hat{m}(\hat{\beta})\right)$,
but we provide its formula to give a clearer definition of $V$.

\begin{proposition}\label{prop:=000020var=000020of=000020moment=000020function}

Let $V$ denote $\operatorname{Var}\left(g_{aipw}\left(O;\beta,\eta(\beta)\right)\right)$
evaluated at $\beta=\beta^{0}$. Under the assumptions SMAR, Overlap,
and M,
\begin{align*}
V & =\Var\left(\E\left[g|X_{1},W\right]\right)+\E\left[\left(\frac{1}{p_{1}}-1\right)\Var\left(g_{1}|W\right)\right]\\
 & \;\;+\E\left[\left(1-p_{1}\right)\left(\frac{1-p_{1}}{p_{01}}-1\right)\Var\left(\mu_{21}|W\right)\right]\\
 & \;\;+\E\left[\left(\frac{p_{1}^{2}}{p_{11}}+\frac{\left(1-p_{1}\right)^{2}}{p_{01}}\right)\Var\left(g_{2}|X_{1},W\right)\right].
\end{align*}
Here, $g\equiv g(X,W;\beta^{0})$, $g_{1}\equiv g_{1}(X_{1},W;\beta^{0})$
, $g_{2}\equiv g_{2}(X_{2},W;\beta^{0})$ and $\mu_{21}\equiv\mu_{21}(X_{1},W;\beta^{0})$. 

\end{proposition}

For the first-step estimation, construction of $\hat{p}$ and $\hat{m}(\beta)$
depends on the researcher's prior beliefs about model structures.
Parametric approaches are natural starting points, especially when
double robustness permits certain forms of misspecification. When
the nuisance structure is unclear, nonparametric approaches are also
natural to avoid overly restrictive constraints. We apply sieve estimation
with series and spline bases for the subsequent simulations and the
empirical application, and provide asymptotic properties for them
later in this section. Technical details on sieve estimation with
series/spline bases can be found in \citet{ai2003efficient} and \citet{chen2007large}.
A more relevant discussion on estimation of nuisance parameters can
be found in Appendix B of \citet{chaudhuri2016gmm}. In practice,
the complexity of the sieve approximation (e.g., truncation number,
order of series, number of knots) should increase with the sample
size and can be selected via cross-validation. 

We want to first show that first-stage estimation does not affect
second-stage asymptotic distribution. The orthogonality result in
Section \ref{subsec:Double-Robustness} does not itself guarantee
this. It implies that the first-order derivative term associated with
nuisance estimation is zero. To ensure the remaining first-step effect
is asymptotically negligible, we impose regularity assumptions for
the first-step estimation.

\begin{asR}\label{assn:=000020R}
\begin{itemize}
\item [(1)] There exists a neighborhood $\mathcal{N}$ of $\beta^{0}$
such that the components of $\hat{\eta}(\beta)$ are measurable functions
of the corresponding information set for $\forall\beta\in\mathcal{N}$.
The components of $\hat{p}$ are bounded away from zero. 
\item [(2)] Let $\hat{h}_{n}=\hat{\eta}(\beta^{0})-\eta^{0}(\beta^{0})=\left(\hat{h}_{p,n},\hat{h}_{m,n}\right).$
There exists a measurable linear map $h\equiv(h_{p},h_{m})\rightarrow D_{i}(h)$
such that 
\[
\hat{g}_{aipw}\left(\beta^{0},\hat{\eta}(\beta^{0})\right)-\frac{1}{n}\sum_{i=1}^{n}g_{aipw}\left(O_{i};\beta^{0},\eta^{0}(\beta^{0})\right)=\frac{1}{n}\sum_{i=1}^{n}D_{i}\left(\hat{h}_{n}\right)+o_{p}(\frac{1}{\sqrt{n}}),
\]

and for every admissible $h_{p}$ and $h_{m}$,
\begin{align*}
\E\left[D_{i}(h_{p},0)\right] & =\frac{d}{dr}\E\left[g_{aipw}\left(O;\beta^{0},(p^{0}+rh_{p},m{}^{0})\right)\right]\vert_{r=0}\\
\E\left[D_{i}(0,h_{m})\right] & =\frac{d}{dr}\E\left[g_{aipw}\left(O;\beta^{0},(p^{0},m^{0}+rh_{m})\right)\right]\vert_{r=0}.
\end{align*}

\item [(3)] For the linearization remainder term,
\[
\frac{1}{\sqrt{n}}\sum_{i=1}^{n}\left[D_{i}\left(\hat{\eta}(\beta^{0})-\eta_{0}\right)-E\left[D\left(\hat{\eta}(\beta^{0})-\eta_{0}\right)\right]\right]\rightarrow_{p}0.
\]
\item [(4)] The nuisance estimators are uniformly consistent on $\mathcal{\mathcal{B}}$,
$\sup_{\beta\in\mathcal{B}}||\hat{\eta}(\beta)-\eta^{0}(\beta)||=o_{p}(1).$
\end{itemize}
\end{asR}

The high-level assumptions in Assumption R can be found in Section
5 of \citet{newey1994asymptotic}. R(1) ensures all the estimators
are admissible in our framework. R(2)-(4) are the linearization, stochastic
equicontinuity and uniform convergence conditions imposed in \citet{newey1994asymptotic}.
There is one more mean-square continuity condition required, but it
automatically holds under Neyman orthogonality stated in Corollary
\ref{lem:=000020neyman-ortho}. The orthogonality guarantees that
the first-order asymptotic distribution is not disturbed by the first-stage
estimation, while Assumption R(2) and R(3) control the second-order
remainder terms. These conditions together provide sufficient conditions
on asymptotic normality of the moment function, and R(4) guarantees
the asymptotic normality of $\hat{\beta}$. These conditions can be
satisfied in standard parametric first-step settings under ordinary
smoothness and root-$n$ convergence. For nonparametric first-stage
estimation, Assumption R can be satisfied with sufficient smoothness
of the nuisance functions and an appropriate growth rate for the parameter
dimension. In the Appendix, we show the relevant conditions for sieve
estimation such that Assumption R holds. 

Under the regularity conditions in Assumption R, disturbance from
the first-stage estimation is excluded and root-$n$ convergence can
be guaranteed.

\begin{theorem}\label{thm:=000020root=000020n}

Suppose Assumptions SMAR, Overlap, M and R hold. Let $G=G(\beta^{0})$,
and $V=\operatorname{V}\left(g_{aipw}(O;\beta^{0},\eta^{0}(\beta^{0}))\right)$.
Let $\hat{\beta}$ denote the final estimator following the proposed
estimation procedure. Then 
\[
\sqrt{n}\left(\hat{\beta}-\beta^{0}\right)\rightarrow_{d}N\left(0,\left(G^{'}V^{-1}G\right)^{-1}\right).
\]

\end{theorem}

\subsection{Semiparametric Efficiency Bound}

Another theoretical contribution of this paper is to show that, under
the proposed assumption, the estimator attains the semiparametric
efficiency bound even when strengthened MAR fails. Achieving a closed-form
semiparametric efficiency bound is difficult under non-monotone missingness.
The complication is that heterogeneous missing patterns can break
the Neyman orthogonality conditions needed for a closed-form influence
function. \citet{chaudhuri2016gmm} propose a setting where all missingness
mechanisms are independent of missing values conditional on the same
set of fully observed variables. This uniform conditioning makes it
easier to control unwanted influence on variation of $\beta$. In
contrast, our framework allows an asymmetric missingness mechanism
such that different missing patterns depend on different conditioning
sets, including some partially observed variables. To control for
unwanted variation and restore pathwise differentiability, we construct
asymmetric AIPW components for $X_{1}$ and $X_{2}$. The additive
separability of the original moment $g=g_{1}+g_{2}$ is crucial. It
lets us enforce Neyman orthogonality separately with respect to the
relevant conditioning sets for $X_{1}$ and $X_{2}$. Intuitively,
the two sources of missingness are isolated into two terms, each orthogonalized
under its own information set, yielding a closed-form influence function
despite the complex missing pattern.

\begin{theorem}\label{thm:=000020efficiency}

Suppose SMAR, Overlap, M hold, the lower bound of the asymptotic variance
of any regular estimator is given by $\Omega=(G'V^{-1}G)^{-1}$. An
estimator with an asymptotic variance $\Omega$ has the following
asymptotic linear representation:
\[
\sqrt{n}(\hat{\beta}-\beta^{0})=\frac{1}{\sqrt{n}}\sum_{i=1}^{n}\psi(O_{i};\beta^{0})+o_{p}(1),
\]
where $\psi(O)=-\Omega G'V^{-1}g_{aipw}(O;\beta^{0},\eta(\beta^{0})).$ 

\end{theorem}

By Theorem \ref{thm:=000020efficiency}, the estimator $\hat{\beta}$
obtained from the three-step GMM estimation satisfies the linear representation
and $\sqrt{n}(\hat{\beta}-\beta^{0})$ attains the semiparametric
efficiency bound.

\section{Simulation\protect\label{sec:Simulation}}

\subsection{Baseline Simulation}

For the simulations, we consider examples where the endogenous treatment
and outcome variables are subject to missingness at two stages. This
is a commonly encountered situation in empirical studies. When the
missing regressor is endogenous, the propensity of observing both
variables is correlated with an endogenous variable, and therefore
a simple CC analysis results in a biased estimator, but this bias
has often been overlooked. We consider the following model:
\begin{align}
X_{1i} & =1(0.1+0.3W_{1i}+0.1W_{2i}\geq u_{i}),\nonumber \\
X_{2i} & =\alpha X_{1i}+\beta W_{2i}+\epsilon_{i}\equiv0.3X_{1i}+0.5W_{2i}+\epsilon_{i},\label{eq:=000020simulation-DGP}
\end{align}
where $X_{1i}$ and $X_{2i}$ are partially observed treatment and
outcome. $W_{i}\equiv(W_{1i},W_{2i})$ are the fully observed variables,
with $W_{1i}$ as an instrument variable. $(\epsilon_{i},u_{i})$
are jointly normally distributed with positive correlation. The missingness
indicators are determined by the following model:
\begin{align}
R_{1i} & =1(0.2+0.2W_{2i}+0.3W_{1i}\geq\nu_{1i}),\nonumber \\
R_{2i} & =1(0.3-0.05W_{2i}+0.2W_{1i}+0.3R_{1i}X_{1i}\geq\nu_{2i}).\label{eq:=000020simulation-missing}
\end{align}

\begin{table}
\caption{Monte Carlo Simulation with Different Values for $corr(\epsilon,u)$}
\label{table:=000020monte=000020carlo=0000201}
\centering{}%
\begin{tabular}{lcccccc}
\hline 
 & \multicolumn{3}{c}{$\alpha=0.3$} & \multicolumn{3}{c}{$\beta=0.5$}\tabularnewline
\hline 
 & $\hat{\alpha}$ & Mean Bias & RMSE & $\hat{\beta}$ & Mean Bias & RMSE\tabularnewline
\hline 
\multicolumn{7}{l}{n = 1000, R = 200, $corr(\epsilon,u)=0.8$}\tabularnewline
Complete Case & 0.1605 & -0.1395 & \textbf{0.1696} & 0.4917 & -0.0083 & 0.1092\tabularnewline
IPW & 0.3005 & 0.0005 & \textbf{0.1072} & 0.4907 & -0.0093 & 0.1396\tabularnewline
AIPW-monotone & 0.2982 & -0.0018  & \textbf{0.1065} & 0.4975 & -0.0018  & 0.1250\tabularnewline
AIPW-CG2016 & 0.3582 & 0.0582 & \textbf{0.1024} & 0.4975 & -0.0025 & 0.0961\tabularnewline
AIPW-this paper & 0.3041 & 0.0041 & \textbf{0.0769} & 0.4966 & -0.0034 & 0.0747\tabularnewline
\hline 
\multicolumn{7}{l}{n = 1000, R = 200, $corr(\epsilon,u)=0.5$}\tabularnewline
Complete Case & 0.2087 & -0.0913 & \textbf{0.1394} & 0.4940 & -0.0060 & 0.1160\tabularnewline
IPW & 0.2974 & -0.0026 & \textbf{0.1199} & 0.4967 & -0.0033 & 0.1468\tabularnewline
AIPW-monotone & 0.2953 & -0.0047 & \textbf{0.1200} & 0.4995 & -0.0005 & 0.1414\tabularnewline
AIPW-CG2016 & 0.3205 & 0.0205 & \textbf{0.0859} & 0.4847 & -0.0153 & 0.0963\tabularnewline
AIPW-this paper & 0.2985 & -0.0015  & \textbf{0.0799} & 0.4932 & -0.0068 & 0.0826\tabularnewline
\hline 
\multicolumn{7}{l}{n = 1000, R = 200, $corr(\epsilon,u)=0.3$}\tabularnewline
Complete Case & 0.2416 & -0.0584 & \textbf{0.1239} & 0.4957 & -0.0043 & 0.1192\tabularnewline
IPW & 0.2958 & -0.0042 & \textbf{0.1256} & 0.5002 & 0.0002 & 0.1490\tabularnewline
AIPW-monotone & 0.2940  & -0.0060 & \textbf{0.1256} & 0.4993 & -0.0007 & 0.1483\tabularnewline
AIPW-CG2016 & 0.2944 & -0.0056 & \textbf{0.0846} & 0.4908 & -0.0092 & 0.0945\tabularnewline
AIPW-this paper & 0.2959 & -0.0041 & \textbf{0.0818} & 0.4912 & -0.0088 & 0.0875\tabularnewline
\hline 
\end{tabular}
\end{table}

Table \ref{table:=000020monte=000020carlo=0000201} shows the simulation
results for different strategies: CC, IPW, the monotone AIPW estimator
that ignores the non-monotone missing component, the AIPW proposed
in \citet{chaudhuri2016gmm} (CG2016), and our AIPW. We also vary
the correlation between $\epsilon$ and $u$ for different endogeneity
levels of $X_{1}$. Because $R_{2}$ depends on $X_{1}$, higher $corr(\epsilon,u)$
implies a more endogenous missingness mechanism, and the CC estimator
becomes more biased. With a positive correlation between $\epsilon$
and $u$, $\epsilon$ is negatively correlated with $X_{1}$, so the
bias has a negative sign. 

While the other IPW and AIPW estimators remain unbiased, the AIPW
approach proposed in this paper has the smallest RMSE across designs.
For both the IPW and AIPW-type estimators, we apply series estimation
with cross-validation to choose polynomial order for the nuisance
functions. The IPW estimators have higher RMSE than the CC estimator
in some cases, likely because it still drops all incomplete observations
yet estimates nuisance parameters nonparametrically with limited data.
We also show results from the other AIPW-type estimators. When we
force a monotone-only AIPW (dropping the last two components in the
augmentation term), we find that its statistical performance is no
better than that of the IPW, highlighting the value of observations
with $R_{1}=0$ and $R_{2}=1$. On the other hand, when we apply an
alternative non-monotone AIPW proposed in \citet{chaudhuri2016gmm},
its efficiency remains lower than that of the proposed estimator,
especially when $R_{2}$ is highly endogenous. 

\subsection{Simulation with Misspecification}

In Table \ref{table:=000020Monte=000020Carlo=0000202}, we provide
examples of two types of misspecification: omitted variables with
a nonparametric estimation strategy, and incorrect model specification
with a parametric estimation strategy. We continue to use the same
DGP with $corr(\epsilon,u)=0.3$ for these exercises. For the first
block, we drop $W_{1}$ from estimating all imputations to create
misspecified imputed values, and we drop $X_{1}$ when estimating
$p_{11}$ to generate misspecified missingness propensities. We still
apply sieve estimation with series basis functions, with cross-validated
polynomial orders. These estimators are unbiased but suffer from slight
efficiency loss compared to the results with both parts correctly
specified in Table \ref{table:=000020monte=000020carlo=0000201}.
For the wrong models, we use parametric models to estimate the nuisance
parameters. We use probit to estimate the propensities as well as
all $\E[X_{1}\mid W]$, and we apply linear models to estimate $\E[X_{2}\mid W],\E[X_{2}\mid X_{1},W]$.
We include unnecessary quadratic terms of $W$ as the misspecification.
These estimators still perform well. The RMSE is lower than the one
in Table \ref{table:=000020monte=000020carlo=0000201}, but it is
driven by a smaller standard deviation when parametric estimation
is applied in finite sample. These results confirm the double robustness
property.

\begin{table}
\caption{Monte Carlo Simulation with Misspecification}
\label{table:=000020Monte=000020Carlo=0000202}
\centering{}%
\begin{tabular}{ccccccc}
\hline 
 & \multicolumn{3}{c}{$\alpha=0.3$} & \multicolumn{3}{c}{$\beta=0.5$}\tabularnewline
\hline 
 & $\hat{\alpha}$ & Mean Bias & RMSE & $\hat{\beta}$ & Mean Bias & RMSE\tabularnewline
\hline 
\multirow{4}{*}{Omitted Variable} & \multicolumn{6}{l}{misspecified imputed values}\tabularnewline
 & 0.3001 & 0.0001 & 0.0908 & 0.5003 & 0.0003 & 0.0886\tabularnewline
 & \multicolumn{6}{l}{misspecified missingness propensities}\tabularnewline
 & 0.3002 & 0.0002 & 0.0907 & 0.5005 & 0.0005 & 0.0886\tabularnewline
\hline 
\multirow{4}{*}{Wrong Model} & \multicolumn{6}{l}{misspecified imputed values}\tabularnewline
 & 0.2955 & -0.0045 & 0.0811 & 0.4928 & -0.0072 & 0.0854\tabularnewline
 & \multicolumn{6}{l}{misspecified missingness propensities}\tabularnewline
 & 0.2957 & -0.0043  & 0.0811 & 0.4930 & -0.0070 & 0.0850\tabularnewline
\hline 
\end{tabular}
\end{table}

\subsection{Simulation with MAR}

As discussed before in Section \ref{subsec:Efficient-Estimator},
the estimator proposed by this paper retains double robustness, and
when all propensities are specified as functions of $W$, it has the
same first-order efficiency interpretation as the moment in \citet{chaudhuri2016gmm},
following their tangent space characterization. In this subsection,
we keep using the main model in \ref{eq:=000020simulation-DGP} with
$\operatorname{corr}(\epsilon,u)=0.3$, but remove the $R_{1}X_{1}$
term from the $R_{2}$ equation in \ref{eq:=000020simulation-missing},
i.e., $R_{2i}=1(0.3-0.05W_{2i}+0.2W_{1i}\geq\nu_{2}).$ 

\begin{table}
\caption{Monte Carlo Simulation when Strengthened MAR Holds}
\label{table:=000020Monte=000020Carlo=0000202-1}
\centering{}%
\begin{tabular}{V{\linewidth}cccccc}
\hline 
 & \multicolumn{3}{c}{$\alpha=0.3$} & \multicolumn{3}{c}{$\beta=0.5$}\tabularnewline
\hline 
 & $\hat{\alpha}$ & Mean Bias & RMSE & $\hat{\beta}$ & Mean Bias & RMSE\tabularnewline
\hline 
\multicolumn{7}{l}{n = 1000, R = 500, $corr(\epsilon,u)=0.3$}\tabularnewline
AIPW-CG & 0.2960 & -0.0040 & 0.0886 & 0.4930 & -0.0070 & 0.0963\tabularnewline
AIPW-this paper

(misspecified propensity) & 0.2966  & -0.0034 & 0.0884 & 0.4919 & -0.0081 & 0.0934\tabularnewline
AIPW-this paper & 0.2959 & -0.0041 & 0.0869 & 0.4925 & -0.0075 & 0.0894\tabularnewline
\hline 
\end{tabular}
\end{table}

The first row presents results from the approach of \citet{chaudhuri2016gmm}.
In the second row, we report results from the exact moment function
$g_{aipw}$ while keeping $p_{11}$ as a function of both $X_{1}$
and $W$. Under a MAR DGP, this is a misspecification of the missingness
propensity, but because double robustness holds under MAR, the resulting
estimator remains valid (discussed previously in Section \ref{subsec:Efficient-Estimator}).
In the last row, we use $g_{aipw}$ with the missing propensities
correctly specified as functions of $W$ only. These estimators' performances
are comparable. Our approach shows a negligible advantage over AIPW-CG,
but we attribute this to the additive separability of the DGP aligning
more closely with our augmentation terms, yielding slightly smaller
standard deviations. This is only a finite-sample effect; asymptotically,
the two estimators attain the same efficiency.

\section{An Empirical Example\protect\label{sec:Application-on-the}}

We apply the AIPW method to the OHIE data, the running example discussed
in previous sections. The AIPW approach is particularly valuable for
improving efficiency in empirical studies with limited sample sizes.
This gain in efficiency directly affects the statistical significance
of parameter estimates. To highlight this improved performance, we
focus on a sample of 3,036 individuals aged 60 and above, evaluating
the effect of OHP enrollment on health-related outcomes among older
adults. 

The variables and notation used here are the same as in the running
example. We denote the fully observed IV, Selected in the lottery,
as $W_{1}$. Other fully observed covariates are denoted as $W_{2}$
and include Number of people in household, Age, Female, Living in
a metropolitan area, and English-speaking. Among these variables,
Number of people in household is a key control variable to ensure
exogeneity of the IV. For this variable, 1 stands for a household
with a single member, while 2 and 3 represent households with two
and more than two members. We keep the notation $X_{1}$ and $X_{2}$
for treatment and outcomes. The treatment is enrollment in OHP, including
both Standard and Plus plans, and outcomes are Physical activities,
Depression, Got all needed medical care and Got all needed dental
care. Physical activities is rated on a three-point scale: 1 (more
active), 2 (same) and 3 (less active). Depression level is constructed
from two questions: ``How often have you been disinterested in doing
things in the past two weeks?'' and ``How often have you felt depressed
in the past two weeks?'' These variables measure health from physical
and mental perspectives, as well as satisfaction with the medical
care and services outside primary care. Summary statistics are presented
in Table \ref{tab: sum}. 

\begin{table}[ht]
\caption{Summary Statistics in the OHIE Data}
\label{tab: sum}
\centering
\resizebox{\textwidth}{!}{
\begin{tabular}{llccccc}
\toprule
Group & Variable & Count & Mean & SD & Min & Max \\
\midrule
$W_1$ & Selected in the lottery & 3036 & 0.5095 & 0.5000 & 0 & 1 \\
\midrule
\multirow{5}{*}{$W_2$}
  & Number of people in household        & 3036 & 1.2567 & 0.4422 & 1 & 3 \\
  & Age                                   & 3036 & 61.3785 & 1.0994 & 60 & 63 \\
  & Female                                & 3036 & 0.6007 & 0.4898 & 0 & 1 \\
  & Zip code in a metropolitan statistical area
                                          & 3036 & 0.7344 & 0.4417 & 0 & 1 \\
  & Requested English-language materials  & 3036 & 0.9147 & 0.2793 & 0 & 1 \\
\midrule
\multirow{5}{*}{$X_2$}
  & Physical activities                   & 1696 & 2.1433 & 0.7628 & 1 & 3 \\
  & Depression                            & 1682 & 1.9792 & 0.9475 & 1 & 4 \\
  & Got all needed medical care           & 1637 & 0.6872 & 0.4638 & 0 & 1 \\
  & Got all needed dental care            & 1670 & 0.4766 & 0.4996 & 0 & 1 \\
  & \multicolumn{5}{l}{} \\[-0.9em] 
\midrule
$X_1$ & Currently have OHP insurance       & 1628 & 0.1210 & 0.3262 & 0 & 1 \\
\bottomrule
\end{tabular}
}
\end{table}

Let $W$ be a vector of the fully observed IV, control variables and
an all-ones vector. The original moment condition comes from a simple
linear model:
\[
X_{2}=\beta_{0}+\beta_{1}X_{1}+W_{2}^{'}\beta_{W}+\epsilon\Rightarrow\E\left[W\left(X_{2}-\beta_{0}-\beta_{1}X_{1}-W_{2}^{'}\beta_{W}\right)\right]=0.
\]
We apply our AIPW approach, estimating the nuisance via sieve with
a B-spline basis. The regression results in Tables \ref{tab: reg1}
and \ref{tab: reg2} show that the AIPW method substantially improves
estimation efficiency. Under CC and IPW, only the Depression coefficient
is significant at the 1\% level, while the AIPW estimators are significant
at the 0.1\% level. For the other outcome variables, CC and IPW estimators
are not significant, while the AIPW estimators are significant at
the 0.1\% level. These findings indicate that the OHP program improves
health-related outcomes among older adults, significantly enhancing
physical activity and reducing depression levels. It also increases
the probability of receiving adequate primary and dental care by 0.268
and 0.296, respectively. Such significant effects can be overlooked
when incomplete observations are dropped.

\begin{sidewaystable} 
\caption{Regression Results: Health Status}
\label{tab: reg1}
\def\sym#1{\ifmmode^{#1}\else\(^{#1}\)\fi}
\begin{tabular}{l*{6}{c}}
\hline\hline                               
&\multicolumn{3}{c}{Physical Activities}&\multicolumn{3}{c}{Depression}\\
\cmidrule{2-4}\cmidrule{5-7}
&\multicolumn{1}{c}{(1)}&\multicolumn{1}{c}{(2)}&\multicolumn{1}{c}{(3)}&\multicolumn{1}{c}{(1)}&\multicolumn{1}{c}{(2)}&\multicolumn{1}{c}{(3)}\\    
&\multicolumn{1}{c}{CC GMM}&\multicolumn{1}{c}{IPW GMM}&\multicolumn{1}{c}{AIPW GMM}&\multicolumn{1}{c}{CC GMM}&\multicolumn{1}{c}{IPW GMM}&\multicolumn{1}{c}{AIPW GMM}\\
&                 &                 &                     &                    &                    &\\ 
OHP                   
&   -0.460        &   -0.457        &   -0.500\sym{***}   &   -0.932\sym{**}   &   -0.912\sym{**}   &   -0.931\sym{***}\\
&  (0.271)        &   (0.270)       &   (0.134)           &   (0.338)          &   (0.335)          &   (0.181)\\
&                 &                 &                     &                    &                    &          \\   
Female
&   0.0540        &   0.0506        &   0.0124            &   -0.00906         &   -0.0132          &   0.0455\\  
&  (0.0471)       &   (0.0471)      &   (0.0229)          &   (0.0581)         &   (0.0581)         &   (0.0313)\\
&                 &                 &                     &                    &                    &            \\ 
Number of Household Members             
&  -0.0786        &  -0.0860        &   -0.0431           &   -0.194\sym{**}   &   -0.201\sym{**}   &   -0.193\sym{***}\\                  
&  (0.0558)       &  (0.0559)       &   (0.0267)          &   (0.0654)         &   (0.0666)         &   (0.0352)\\
&                 &                 &                     &                    &                    &            \\ 
Age             
&  -0.0314        &  -0.0333        &   -0.0340\sym{***}  &  -0.0601\sym{*}    &  -0.0634\sym{*}    &   -0.101\sym{***}\\  
&  (0.206)        &  (0.0206)       &   (0.0101)          &  (0.0257)          &  (0.0257)          &   (0.0133)\\
&                 &                 &                     &                    &                    &            \\ 
MSA             
&   0.0335        &   0.0275        &    0.00230          &  -0.0197           &   -0.0241          &    0.0415\\
&  (0.0492)       &  (0.0492)       &    (0.0234)         &  (0.0623)          &   (0.0625)         &    (0.0319)\\
&                 &                 &                     &                    &                    &            \\
English-Speaking              
&   0.126         &   0.118         &    0.0150           &   -0.00293         &   -0.0183          &-0.175\sym{*}\\                     
&  (0.102)        &  (0.102)        &   (0.0498)          &   (0.123)          &   (0.125)          &(0.0702)\\
&                 &                 &                     &                    &                    &         \\
Constant             
&  4.026\sym{**}  &  4.169\sym{**}  &  4.337\sym{***}     &  6.012\sym{***}    &  6.237\sym{***}    &  8.642\sym{***}\\                     
&  (1.301)        &  (1.300)        &  (0.627)            &  (1.616)           &  (1.619)           &  (0.836)\\ 
\hline
Observations        
&       1222      &  1220           &   3036              &   1215             &  1213              &   3036\\ 
\hline\hline 
\multicolumn{7}{l}{\footnotesize Standard errors in parentheses}\\ 
\multicolumn{7}{l}{\footnotesize \sym{*} \(p<0.05\), \sym{**} \(p<0.01\), \sym{***} \(p<0.001\).}\\ 
\end{tabular}
\end{sidewaystable}    

\begin{sidewaystable} 
\caption{Regression Results: Medical Service}
\label{tab: reg2}
\def\sym#1{\ifmmode^{#1}\else\(^{#1}\)\fi}
\begin{tabular}{l*{6}{c}}
\hline\hline                               
&\multicolumn{3}{c}{Got All Needed Medical Care}&\multicolumn{3}{c}{Got All Needed Dental Care}\\
\cmidrule{2-4}\cmidrule{5-7}
&\multicolumn{1}{c}{(1)}&\multicolumn{1}{c}{(2)}&\multicolumn{1}{c}{(3)}&\multicolumn{1}{c}{(1)}&\multicolumn{1}{c}{(2)}&\multicolumn{1}{c}{(3)}\\    
&\multicolumn{1}{c}{CC GMM}&\multicolumn{1}{c}{IPW GMM}&\multicolumn{1}{c}{AIPW GMM}&\multicolumn{1}{c}{CC GMM}&\multicolumn{1}{c}{IPW GMM}&\multicolumn{1}{c}{AIPW GMM}\\
&                 &                 &                     &                    &                    &\\ 
OHP                     
&   0.268         &   0.264         &   0.268\sym{***}    &    0.224           &    0.238           &    0.296\sym{***}\\     
&  (0.156)        &   (0.158)       &   (0.0814)          &   (0.169)          &   (0.174)          &   (0.0882)\\
&                 &                 &                     &                    &                    &           \\   
Female
&  -0.0398        &  -0.0408        &  -0.0245            &    0.000966        &   0.000643         &  -0.0290\\  
&  (0.0279)       &   (0.0280)      &   (0.0139)          &   (0.0305)         &   (0.0306)         &   (0.0151)\\
&                 &                 &                     &                    &                    & \\ 
Number of Household Members             
&   0.0735\sym{*} &   0.0717\sym{*} &    0.121\sym{***}   &    0.0916\sym{**}  &    0.0944\sym{**}  &    0.0729\sym{***}        \\                  
&  (0.0306)       &  (0.0319)       &   (0.0156)          &   (0.0352)         &   (0.0356)         &   (0.0175)\\
&                 &                 &                     &                    &                    &           \\ 
Age             
&   0.0217        &   0.0217        &    0.0303\sym{***}  &   0.0214           &   0.0227           &    0.0235\sym{***}\\  
&  (0.0126)       &  (0.0127)       &   (0.00611)         &  (0.0133)          &  (0.0133)          &   (0.00657)\\
&                 &                 &                     &                    &                    &             \\ 
MSA             
&   0.0332        &   0.0349        &    0.0193           &   0.00802          &    0.00894         &   -0.0103 \\
&  (0.0300)       &  (0.0302)       &    (0.0147)         &  (0.0322)          &   (0.0323)         &    (0.0157)\\
&                 &                 &                     &                    &                    &             \\
English-Speaking              
&   0.0215        &   0.0223        &    0.00684          &   -0.187\sym{**}   &   -0.182\sym{**}   &-0.115\sym{***} \\                     
&  (0.0636)       &  (0.0640)       &   (0.0312)          &   (0.0638)         &   (0.0643)         &(0.0340)\\
&                 &                 &                     &                    &                    &         \\
Constant             
& -0.777          &  -0.775         &  -1.374\sym{***}    &  -0.814            &  -0.902            &  -0.962\sym{*} \\                     
&  (0.795)        &  (0.801)        &  (0.382)            &  (0.835)           &  (0.836)           &  (0.411)\\ 
\hline
Observations        
&       1177      &  1175           &   3036              &   1203             &  1201              &   3036\\  
\hline\hline 
\multicolumn{7}{l}{\footnotesize Standard errors in parentheses}\\ 
\multicolumn{7}{l}{\footnotesize \sym{*} \(p<0.05\), \sym{**} \(p<0.01\), \sym{***} \(p<0.001\).}\\ 
\end{tabular}
\end{sidewaystable}    

In this empirical analysis, we do not find evidence that the CC estimator
differs substantially from the AIPW estimator due to bias. One possible
explanation is that the dependence between $R_{2}$ and $X_{1}$ may
be weaker than in the simulations. An interesting finding is that
for some outcomes, such as ``Depression,'' the estimates from the
CC estimator are closer to those from the AIPW approach than the IPW
estimates. Although these differences are negligible and tests suggest
equivalence among all estimators, the small deviations may be attributable
to the inclusion of estimated nuisance parameters. However, the AIPW
estimator retains its double robustness property and is therefore
less sensitive to such nuisance-parameter estimation. 

\section{Conclusion\protect\label{sec:Conclusion}}

This paper addresses the issue of non-monotone missingness at two
stages and develops an estimator that is both unbiased and efficient.
Under the SMAR assumption, we derive a closed-form efficient influence
function, which allows us to propose an AIPW estimator that achieves
the corresponding efficiency bound. The AIPW estimator has an asymmetric
form for missingness at different stages, thereby retaining double
robustness even when the missingness mechanisms depend on different
sets of variables. 

We provide an empirical example to support the SMAR assumption. In
this example, when the first-stage treatment variable is observed,
it is associated with the missingness mechanism of outcome variables
collected at the second stage. Conversely, when the first-stage treatment
variable is not observed, there is no evidence of correlation between
the second-stage missingness mechanism and a proxy for the first-stage
treatment. No analogous asymmetric dependence is found between the
first-stage missingness mechanism and the second-stage partially observed
variables. In the empirical example, the AIPW approach reduces the
standard errors by approximately 50\% across all four outcomes compared
to the commonly used CC estimator. This improvement directly changes
the statistical significance of the estimated treatment effects, indicating
significant beneficial effects on health-related outcomes among older
adults. Simulation exercises further support the desirable properties
of the AIPW estimator, confirming its efficiency and robustness in
practical applications.

\begin{appendix}

\section{Sieve First Stage Estimation}

The conditions in this section are mostly adapted from well-established
sieve regularity conditions in the literature (\citet{newey1994asymptotic};
\citet{chen2008semiparametric}; \citet{cattaneo2010efficient}; \citet{chaudhuri2016gmm}),
and closely follow those used in  \citet{cattaneo2010efficient} and
\citet{chaudhuri2016gmm}. They ensure that the estimated nuisance
parameters converge fast enough so as not to affect the convergence
rate of the second-step estimation. 

\begin{corollary}\label{cor:=000020sieve=000020estimates}

For each primitive nuisance component $e\in\mathcal{E}$, let $Z_{e}$
denote its argument vector with support $\mathcal{Z}_{e}$, and $\mathcal{E}$
contains all primitive propensity and imputation components estimated
in the first step. Suppose:
\begin{itemize}
\item [(1)] For each $e$, $\mathcal{Z}_{e}$ is compact and its density
is bounded away from zero and infinity on the support. The propensity
score estimators are bounded away from 0.
\item [(2)] The true function $e^{0}$ belongs to a H\"{o}lder class of
smoothness $s_{e}>0$, where $d_{e}$ denotes the dimension of $Z_{e}$.
\item [(3)] Each nuisance parameter is estimated by sieve with either series
bases or spline bases with $K_{n}$ basis terms. Let $\kappa=1$ for
power-series bases and $\kappa=\frac{1}{2}$ for the spline bases,
and define $\tau=\min_{e\in\mathcal{E}}\frac{s_{e}}{d_{e}}$.
\item [(4)] The basis satisfies the usual sup-norm and Gram-matrix regularity
conditions:
\[
\sup_{z}||b_{K_{n}}(z)||\lesssim K_{n}^{\kappa},\;\;\lambda_{min}\left(\E\left[b_{K_{n}}(Z)b_{K_{n}}(Z)^{'}\right]\right)\geq c>0,
\]
for sufficiently large $n$.
\item [(5)] The sieve approximation error satisfies
\[
\inf_{\alpha}\sup_{z}|e^{0}(z)-b_{K_{n}}(z)^{'}\alpha|\lesssim K_{n}^{-s_{e}/d_{e}}
\]
uniformly over the primitive nuisance components, and the same bound
holds for the $\beta$-derivatives of those imputations that depend
on $\beta$.
\item [(6)] The number of sieve terms satisfies
\[
K_{n}=n^{\nu},\;\;\tau>\frac{5}{2}\kappa+\frac{1}{2},\;\;\frac{1}{4\tau-6\kappa}<\nu<\frac{1}{4\kappa+2}.
\]
\end{itemize}
Under these conditions, the high-level assumptions in Assumption R
are satisfied and consequently 
\[
\sqrt{n}\left(\hat{\beta}-\beta^{0}\right)\rightarrow_{d}N\left(0,\left(G^{'}V^{-1}G\right)^{-1}\right).
\]

\end{corollary}

Proof of this corollary is presented in Appendix \ref{cor:=000020sieve=000020estimates}.

\section{Strengthened MAR}

We next show that the proposed moment function also has desirable
properties under the strengthened MAR benchmark. Suppose the strengthened
MAR holds such that
\[
(R_{1},R_{2})\perp(X_{1},X_{2})|W.
\]
Under this assumption, the true missingness propensities are functions
of $W$ only:
\begin{align*}
p_{1}(W) & =\Pr[R_{1}=1|W],\\
p_{01}(W) & =\Pr[R_{1}=0,R_{2}=1|W],\\
p_{11}(W) & =\Pr[R_{1}=1,R_{2}=1|W].
\end{align*}
We maintain the abbreviated notations $p_{1},p_{01},p_{11}$ for these
$W$-only propensities and impose the analogous overlap conditions
so that all probabilities appearing in denominators are bounded away
from $0$. Neyman orthogonality still holds. The centering conditions
become even simpler because all inverse-probabilities residuals can
be evaluated by conditioning on $W$. Therefore, the AIPW moment remains
correctly centered at $\beta^{0}$ whenever either the propensity
block or the imputation block is correctly specified. Neyman orthogonality
follows as the local implication of these blockwise double-robust
identities.

It is useful to distinguish population validity from the efficiency
comparison. If $p_{11}$ is estimated as a flexible function of $(X_{1},W)$,
the population centering and double-robust identities are not affected
when the relevant imputation block is correctly specified. However,
it may lose efficiency by using a subsample. Therefore, for the strengthened
MAR efficiency comparison, we use the correctly restricted propensities
as functions of $W$.

The result below takes the strengthened MAR tangent-space characterization
of \citet{chaudhuri2016gmm} as the benchmark. Under the same benchmark
representation, the proposed $g_{aipw}$ satisfies the same first-order
efficiency verification. This comparison does not require the proposed
AIPW score and the CG score to be algebraically identical, but their
projections onto the benchmark tangent space coincide.

\begin{proposition}\label{prop:=000020MAR-equiv}

Suppose the strengthened MAR condition holds such that $(R_{1},R_{2})\perp(X_{1},X_{2})|W$.
Assume Overlap and Assumption M. Let all propensities be specified
as functions of $W$, and let $T_{CG}$ denote the tangent space representation
used in \citet{chaudhuri2016gmm} for the strengthened-MAR model.
Let
\[
\psi_{aipw}(O)=-(G^{'}V^{-1}G)^{-1}G^{'}V^{-1}g_{aipw}(O;\beta^{0},\eta^{0}).
\]
Then $g_{aipw}\left(O;\beta^{0},\eta^{0}\right)$ is correctly centered
and satisfies the same pathwise differentiability equations. Following
the same argument in \citet{chaudhuri2016gmm}, $\psi_{aipw}$ has
the same first-order efficiency interpretation as the benchmark case.

\end{proposition}

Equivalently, the influence function constructed from $g_{aipw}$
coincides with the efficient influence function in the model of \citet{chaudhuri2016gmm}.
One advantage is that when the wrong missingness assumption is made,
we can still retain a robust estimator.

\section{Proofs}

\subsection{Proof of Proposition \ref{prop:=000020propensity=000020identification}}

By the chain rule, $\Pr\left[R_{1}=a,R_{2}=b\mid X_{1},X_{2},W\right]=\Pr[R_{2}=b\mid R_{1}=a,X_{1},X_{2},W]\Pr[R_{1}=a\mid X_{1},X_{2},W]$.
For $(R_{1},R_{2})=(0,0)$,
\begin{align*}
\Pr\left[R_{1}=0,R_{2}=0\mid X_{1},X_{2},W\right] & =\Pr[R_{2}=0\mid R_{1}=0,X_{1},X_{2},W]\Pr[R_{1}=0\mid X_{1},X_{2},W]\\
 & =\Pr[R_{2}=0\mid R_{1}=0,W]\Pr[R_{1}=0\mid W]\\
 & =\left(1-p_{1|0}(W)\right)\cdot\left(1-p_{1}(W)\right).
\end{align*}
Similarly, for $(R_{1},R_{2})=(0,1)$,
\begin{align*}
\Pr\left[R_{1}=0,R_{2}=1\mid X_{1},X_{2},W\right] & =\Pr[R_{2}=1\mid R_{1}=0,X_{1},X_{2},W]\Pr[R_{1}=0\mid X_{1},X_{2},W]\\
 & =\Pr[R_{2}=1\mid R_{1}=0,W]\Pr[R_{1}=0\mid W]\\
 & =p_{1|0}(W)\cdot\left(1-p_{1}(W)\right).
\end{align*}
The second equality in each derivation follows from SMAR such that
$R_{2}\perp(X_{1},X_{2})\mid W,R_{1}=0$ and $R_{1}\perp(X_{1},X_{2})\mid W$.
For $(R_{1},R_{2})=(1,0)$ and $(1,1)$,
\begin{align*}
\Pr\left[R_{1}=1,R_{2}=0\mid X_{1},X_{2},W\right] & =\Pr[R_{2}=0\mid R_{1}=1,X_{1},X_{2},W]\Pr[R_{1}=1\mid X_{1},X_{2},W]\\
 & =\Pr[R_{2}=0\mid R_{1}=1,X_{1},W]\Pr[R_{1}=1\mid W]\\
 & =\left(1-p_{1|1}(X_{1},W)\right)\cdot p_{1}(W),
\end{align*}
\begin{align*}
\Pr\left[R_{1}=1,R_{2}=1\mid X_{1},X_{2},W\right] & =\Pr[R_{2}=1\mid R_{1}=1,X_{1},X_{2},W]\Pr[R_{1}=1\mid X_{1},X_{2},W]\\
 & =\Pr[R_{2}=1\mid R_{1}=1,X_{1},W]\Pr[R_{1}=1\mid W]\\
 & =p_{1|1}(X_{1},W)\cdot p_{1}(W).
\end{align*}
where we use $R_{2}\perp X_{2}\mid W,X_{1},R_{1}=1$ and $R_{1}\perp(X_{1},X_{2})\mid W$. 
\begin{flushright}
$\square$
\par\end{flushright}

\subsection{Proof of Proposition \ref{prop:=000020augementation=000020identification}}

By Assumption SMAR, $\E\left[g_{1}(X_{1},W;\beta)\mid W,R_{1}=1\right]=\E\left[g_{1}(X_{1},W;\beta)\mid W\right]=\mu_{1}(W)$,
following from $R_{1}\perp(X_{1},X_{2})|W$. Similarly, using $R_{2}\perp(X_{1},X_{2})\mid W,R_{1}=0$,
and $R_{1}\perp(X_{1},X_{2})\mid W$ (in particular, $R_{1}\perp X_{2}\mid W$),
\begin{align*}
\E\left[g_{2}(X_{2},W;\beta)\mid W,R_{1}=0,R_{2}=1\right] & =\E\left[g_{2}(X_{2},W;\beta)\mid W,R_{1}=0\right]\\
 & =\E\left[g_{2}(X_{2},W;\beta)\mid W\right]=\mu_{20}(W).
\end{align*}
Finally, for the case $R_{1}=R_{2}=1$,
\begin{align*}
\E\left[g_{2}(X_{2},W;\beta)\mid X_{1},W,R_{1}=1,R_{2}=1\right] & =\E\left[g_{2}(X_{2},W;\beta)\mid X_{1},W,R_{1}=1\right]\\
 & =\E\left[g_{2}(X_{2},W;\beta)\mid X_{1},W\right]=\mu_{21}(X_{1},W),
\end{align*}
where the first equality follows from the assumption $R_{2}\perp X_{2}\mid X_{1},W,R_{1}=1$,
and the second equality follows the derivation below under $R_{1}\perp(X_{1},X_{2})\mid W$:
\begin{align*}
\E\left[g_{2}(X_{2},W;\beta)|X_{1},W\right] & =\int g_{2}(x_{2},w;\beta)f_{X_{2}|X_{1},W}(x_{2}\mid x_{1},w)dx_{2}\\
 & =\int g_{2}(x_{2},w;\beta)\frac{f_{X_{1},X_{2}|W}(x_{1},x_{2}\mid w)}{f_{X_{1}|W}(x_{1}|w)}dx_{2}\\
 & =\int g_{2}(x_{2},w;\beta)\frac{f_{X_{1},X_{2}|W}(x_{1},x_{2}\mid w,R_{1}=1)}{f_{X_{1}|W}(x_{1}|w,R_{1}=1)}dx_{2}\\
 & =E\left[g_{2}(X_{2},W;\beta)\mid X_{1},W,R_{1}=1\right].
\end{align*}
After identifying these terms, $\E\left[g(X,W;\beta)|W\right]$ can
be calculated by the law of iterated expectation.
\begin{flushright}
$\square$
\par\end{flushright}

\subsection{Proof of Theorem \ref{thm:=000020identification}}

To prove the theorem, we first show that $\E[g_{aipw}(O;\beta,\eta_{0}(\beta))]=\E[g(X,W;\beta)],\forall\beta\in\mathcal{B}$.
We then proceed with the proof by examining each term individually.
For any fixed $\beta$, we show that
\[
\E\left[\frac{R_{1}R_{2}}{p_{11}}g(X,W;\beta)\right]=E\left[g(X,W;\beta)\right].
\]
By the law of iterated expectation,
\begin{align*}
\E\left[\frac{R_{1}R_{2}}{p_{11}}g(X,W;\beta)\right] & =\E\left\{ \E\left[\frac{R_{1}R_{2}}{p_{11}}g(X,W;\beta)\mid X,W\right]\right\} \\
 & =\E\left\{ g(X,W;\beta)\frac{1}{p_{11}(X_{1},W)}\Pr\left[R_{1}=1,R_{2}=1\mid X,W\right]\right\} \\
 & =\E\left\{ g(X,W;\beta)\frac{p_{1|1}(X_{1},W)\cdot p_{1}(W)}{p_{11}(X_{1},W)}\right\} =\E\left[g(X,W;\beta)\right].
\end{align*}
The third equality follows from the SMAR assumption. Next, we show
that the augmentation terms have mean zero. By Proposition \ref{prop:=000020propensity=000020identification},
\begin{align*}
 & \E\left[1-\frac{R_{1}R_{2}}{p_{11}}\mid X,W\right]=\E\left[\frac{(1-R_{1})R_{2}}{p_{01}}-\frac{R_{1}R_{2}}{p_{11}}\mid X,W\right]=\E\left[\frac{R_{1}}{p_{1}}-\frac{R_{1}R_{2}}{p_{11}}\mid X,W\right]=0.
\end{align*}

By the law of iterated expectation
\begin{align*}
\E\left[1-\frac{R_{1}R_{2}}{p_{11}}\mid X,W\right]=0 & \Rightarrow\E\left[(1-\frac{R_{1}R_{2}}{p_{11}})\mu(W;\beta)\right]=\E\left[\mu(W;\beta)\cdot\E\left[1-\frac{R_{1}R_{2}}{p_{11}}\mid X,W\right]\right]=0.
\end{align*}
Similarly,
\begin{align*}
\E\left[\frac{R_{1}}{p_{1}}-\frac{R_{1}R_{2}}{p_{11}}\mid X,W\right]=0 & \Rightarrow\E\left[\left(\frac{R_{1}}{p_{1}}-\frac{R_{1}R_{2}}{p_{11}}\right)\left(g_{1}(X_{1},W;\beta)-\mu_{1}(W;\beta)\right)\right]\\
 & \;\;=\E\left[\left(g_{1}(X_{1},W;\beta)-\mu_{1}(W;\beta)\right)\E[\frac{R_{1}}{p_{1}}-\frac{R_{1}R_{2}}{p_{11}}\mid X,W]\right]=0,
\end{align*}
and analogously $\E\left[p_{1}\left(\frac{R_{1}}{p_{1}}-\frac{R_{1}R_{2}}{p_{11}}\right)\left(\mu_{21}(X_{1},W;\beta)-\mu_{20}(W;\beta)\right)\right]=0$.
\begin{align*}
 & \E\left[\frac{(1-R_{1})R_{2}}{p_{01}}-\frac{R_{1}R_{2}}{p_{11}}\mid X,W\right]=0\\
 & \Rightarrow\E\left[(1-p_{1})\left(\frac{(1-R_{1})R_{2}}{p_{01}}-\frac{R_{1}R_{2}}{p_{11}}\right)\left(g_{2}(X_{2},W;\beta)-\mu_{20}(W;\beta)\right)\right]\\
 & \;\;=\E\left[(1-p_{1})\left(g_{2}(X_{2},W;\beta)-\mu_{20}(W;\beta)\right)\E\left[\frac{(1-R_{1})R_{2}}{p_{01}}-\frac{R_{1}R_{2}}{p_{11}}\mid X,W\right]\right]=0.
\end{align*}
In conclusion, we prove that $\E[g_{aipw}(O;\beta,\eta_{0}(\beta))]=\E[g(X,W;\beta)],\forall\beta\in\mathcal{B}$.
Since $\E\left[g(X,W;\beta)\right]=0$ if and only if $\beta=\beta^{0}$,
$\E[g_{aipw}(\beta)]=0$ if and only if $\beta=\beta^{0}$.
\begin{flushright}
$\square$
\par\end{flushright}

\subsection{Proof of Theorem \ref{thm:=000020robustness=000020SMAR}\protect\label{subsec:Proof-of-Double-Robustness}}

Theorem \ref{thm:=000020identification} proves the result under condition
(a). We now show the statement also holds with condition (b). 

We first rewrite the AIPW moment as
\begin{align*}
g_{aipw}(O;\beta,\eta) & =\underbrace{\frac{R_{1}R_{2}}{p_{11}}g(X,W;\beta)}_{(1)}+\underbrace{\left(1-\frac{R_{1}R_{2}}{p_{11}}\right)\mu(W;\beta)}_{(2)}\\
 & \;\;+\underbrace{\left(\frac{R_{1}}{p_{1}}-\frac{R_{1}R_{2}}{p_{11}}\right)\left(\mu(X_{1},W;\beta)-\mu(W;\beta)\right)}_{(3)}\\
 & \;\;+\left(1-p_{1}\right)\left\{ \left(\frac{\left(1-R_{1}\right)R_{2}}{p_{01}}-\frac{R_{1}R_{2}}{p_{11}}\right)\left(g_{2}(X_{2},W;\beta)-\mu_{20}(W;\beta)\right)\right.\\
 & \;\;\underbrace{\left.-\left(\frac{R_{1}}{p_{1}}-\frac{R_{1}R_{2}}{p_{11}}\right)\left(\mu_{21}(X_{1},W;\beta)-\mu_{20}(W;\beta)\right)\right\} }_{(4)}.
\end{align*}
The sum of the first three terms can be written as
\begin{align*}
(1)+(2)+(3) & =\mu(W;\beta)+\frac{R_{1}R_{2}}{p_{11}}\left(g(X,W;\beta)-\mu(X_{1},W;\beta)\right)+\frac{R_{1}}{p_{1}}\left(\mu(X_{1},W;\beta)-\mu(W;\beta)\right).
\end{align*}
When $\mu(W;\beta)$ and $\mu(X_{1},W;\beta)$ are correctly specified,
by the moment condition,
\[
\E\left[\mu(W;\beta)\right]=\E\left[g(X,W;\beta)\right].
\]
Moreover, suppose $p_{11}$ is an abbreviation for $p_{11}(X_{1},W)$,
or a function of variables included in $(X_{1},W)$, then
\begin{align*}
 & \E\left[\frac{R_{1}R_{2}}{p_{11}}\left(g(X,W;\beta)-\mu(X_{1},W;\beta)\right)\right]\\
 & =\E\left\{ \E\left[\frac{R_{1}R_{2}}{p_{11}}\left(g(X,W;\beta)-\mu(X_{1},W;\beta)\right)\mid X_{1},W\right]\right\} \\
 & =\E\left\{ E\left[\frac{R_{1}R_{2}}{p_{11}}|X_{1},W\right]\E\left[g(X,W;\beta)-\mu(X_{1},W;\beta)\mid X_{1},W\right]\right\} =0,
\end{align*}
where the second equality holds because under SMAR, $R_{1}R_{2}\perp X_{2}\mid X_{1},W$.
Similarly,
\[
\E\left[\frac{R_{1}}{p_{1}}\left(\mu(X_{1},W;\beta)-\mu(W;\beta)\right)\right]=0,
\]
if $p_{1}$ is specified as a function of $W$ or any variable in
$W$. The last term in the above expression of $g_{aipw}(O;\beta,\eta)$
can be rewritten as
\begin{align*}
(4) & =-(1-p_{1})\frac{R_{1}}{p_{1}}\left(\mu_{21}(X_{1},W;\beta)-\mu_{20}(W;\beta)\right)\\
 & \;\;-(1-p_{1})\frac{R_{1}R_{2}}{p_{11}}\left(g_{2}\left(X_{2},W;\beta\right)-\mu_{21}(X_{1},W;\beta)\right)\\
 & \;\;+(1-p_{1})\frac{(1-R_{1})R_{2}}{p_{01}}\left(g_{2}\left(X_{2},W;\beta\right)-\mu_{20}(W;\beta)\right).
\end{align*}
By analogous arguments, under assumption SMAR, $R_{1}\perp X_{1}\mid W$,
$R_{1}R_{2}\perp X_{2}\mid(X_{1},W)$, and $(1-R_{1})R_{2}\perp X_{2}\mid W$.
When $\mu_{21}(X_{1},W;\beta)$ and $\mu_{20}(W;\beta)$ are correctly
specified, and $p_{1}$, $p_{11}$ and $p_{01}$ are specified on
corresponding information sets, according to the law of iterated expectation,
each term in $(4)$ has mean $0$. Therefore, as long as the imputation
terms are correctly specified and the propensities are specified as
functions of the corresponding information sets, $\E\left[g_{aipw}(X,W;\beta)\right]=\E\left[g(X,W;\beta)\right]$
holds for $\forall\beta\in\mathcal{B}$, and Theorem \ref{thm:=000020identification}
holds.

\begin{flushright}
$\square$
\par\end{flushright}

\subsection{Proof of Proposition \ref{prop:=000020var=000020of=000020moment=000020function}}

The variance of $g_{aipw}$ can be easily derived from its decomposition.
But to derive a cleaner expression, we first reconstruct the moment
function into terms that separate the orthogonal components. Define
\[
A_{1}=g_{1}-\mu_{1},\;\;A_{2}=\mu_{21}-\mu_{20},\;\;A_{3}=g_{2}-\mu_{21}.
\]
Then
\begin{equation}
g=\mu+A_{1}+A_{2}+A_{3}.\label{eq:=000020g-decomposition}
\end{equation}
This equality holds from the fact that 
\begin{align*}
g & =\mu+\left(g-\mu\right)\\
 & =\mu+\left(g_{1}-\mu_{1}\right)+\left(g_{2}-\mu_{20}\right)\\
 & =\mu+\left(g_{1}-\mu_{1}\right)+\left(g_{2}-\mu_{21}\right)+\left(\mu_{21}-\mu_{20}\right).
\end{align*}
Starting from the definition of $g_{aipw}$ in \ref{eq:=000020aipw=000020moment=000020function}
and substituting $g$ with the expression in equation \ref{eq:=000020g-decomposition},
we get
\[
g_{aipw}=\mu+\omega_{1}A_{1}+\omega_{2}A_{2}+\omega_{3}A_{3},
\]
where
\begin{align}
\omega_{1} & =\frac{R_{1}}{p_{1}},\nonumber \\
\omega_{2} & =R_{1}+\left(1-p_{1}\right)\frac{\left(1-R_{1}\right)R_{2}}{p_{01}},\nonumber \\
\omega_{3} & =p_{1}\frac{R_{1}R_{2}}{p_{11}}+\left(1-p_{1}\right)\frac{\left(1-R_{1}\right)R_{2}}{p_{01}}.\label{eq:=000020omega}
\end{align}
Then we derive the variance $V$ by
\[
V=\Var\left(\E\left[g_{aipw}|X_{1},W\right]\right)+\E\left[\Var\left(g_{aipw}|X_{1},W\right)\right].
\]
We next compute the components one by one. We first derive $\E\left[g_{aipw}|X_{1},W\right]$:
\[
\E\left[g_{aipw}|X_{1},W\right]=\mu+A_{1}+A_{2}=\E\left[g|X_{1},W\right].
\]
Next we derive $\Var\left(g_{aipw}|X_{1},W\right)$,
\begin{align*}
\Var\left(g_{aipw}|X_{1},W\right) & =\Var\left(\omega_{1}|X_{1},W\right)A_{1}A_{1}^{'}+\Var\left(\omega_{2}|X_{1},W\right)A_{2}A_{2}^{'}\\
 & \;\;+\E\left[\omega_{3}^{2}|X_{1},W\right]\Var\left(A_{3}|X_{1},W\right).
\end{align*}
Substituting the moment expressions of $\omega_{1}$, $\omega_{2}$
and $\omega_{3}$ in \ref{eq:=000020omega},
\begin{align*}
\Var\left(g_{aipw}|X_{1},W\right) & =\left(\frac{1}{p_{1}}-1\right)A_{1}A_{1}^{'}+\left(1-p_{1}\right)\left(\frac{1-p_{1}}{p_{01}}-1\right)A_{2}A_{2}^{'}\\
 & \;\;+\left(\frac{p_{1}^{2}}{p_{11}}+\frac{\left(1-p_{1}\right)^{2}}{p_{01}}\right)\Var\left(g_{2}|X_{1},W\right),
\end{align*}
because $\Var\left(A_{3}|X_{1},W\right)=\Var\left(g_{2}|X_{1},W\right)$.

Finally, we can derive
\[
\E\left[A_{1}A_{1}^{'}|W\right]=\Var\left(g_{1}|W\right),\;\;\E\left[A_{2}A_{2}^{'}|W\right]=\Var\left(\mu_{21}|W\right).
\]
We obtain that
\begin{align*}
V & =\Var\left(\E\left[g|X_{1},W\right]\right)+\E\left[\left(\frac{1}{p_{1}}-1\right)\Var\left(g_{1}|W\right)\right]\\
 & \;\;+\E\left[\left(1-p_{1}\right)\left(\frac{1-p_{1}}{p_{01}}-1\right)\Var\left(\mu_{21}|W\right)\right]\\
 & \;\;+\E\left[\left(\frac{p_{1}^{2}}{p_{11}}+\frac{\left(1-p_{1}\right)^{2}}{p_{01}}\right)\Var\left(g_{2}|X_{1},W\right)\right].
\end{align*}

\selectlanguage{english}%
\begin{flushright}
$\square$
\par\end{flushright}

\subsection{Proof of Theorem \ref{thm:=000020root=000020n}}

For notational convenience, define 
\[
\overline{g}(\beta)=\E\left[g_{aipw}\left(O;\beta,\eta_{0}(\beta)\right)\right],\;\;\hat{g}_{n}(\beta)=\frac{1}{n}\sum_{i=1}^{n}g_{aipw}\left(O_{i};\beta,\hat{\eta}(\beta)\right),
\]
and write $g_{i}(\beta^{0})$ as abbreviation for $g_{aipw}\left(O_{i};\beta^{0},\eta^{0}(\beta^{0})\right)$.
First we prove the consistency of the preliminary and final GMM estimators.
Define the population criteria 
\[
Q_{0}(\beta)=\overline{g}(\beta)^{'}\Lambda^{0}\overline{g}(\beta),\;\;Q^{*}(\beta)=\overline{g}(\beta)^{'}\Lambda^{*}\overline{g}(\beta).
\]
Since $\Lambda^{0}$ and $V^{-1}$ are symmetric positive definite,
both $Q_{0}(\beta)$ and $Q^{*}(\beta)$ are nonnegative, and 
\[
Q_{0}(\beta)=0\Longleftrightarrow\overline{g}(\beta)=0\Longleftrightarrow\beta=\beta^{0}.
\]
Similarly, $Q^{*}(\beta)=0\Longleftrightarrow\beta=\beta^{0}.$ Thus
$\beta^{0}$ is the unique minimizer of both population criteria. 

Next, define the sample criteria
\[
\hat{Q}_{0}(\beta)=\hat{g}(\beta)^{'}\hat{\Lambda^{0}}\hat{g}(\beta),\;\;Q^{*}(\beta)=\hat{g}(\beta)^{'}\hat{\Lambda^{*}}\hat{g}(\beta).
\]
By Assumption M (2), $\sup_{\beta\in\mathcal{B}}||\overline{g}(\beta)||<\infty$.
By Assumptions M(2) and R(4), uniform consistency of the sample moment
is satisfied such that $\sup_{\beta\in\mathcal{B}}||\hat{g}_{n}(\beta)-\overline{g}(\beta)||=o_{p}(1)$.
Therefore,
\[
\sup_{\beta\in\mathcal{B}}||\hat{g}_{n}(\beta)||\leq\sup_{\beta\in\mathcal{B}}||\hat{g}_{n}(\beta)-\overline{g}(\beta)||+\sup_{\beta\in\mathcal{B}}||\overline{g}(\beta)||=O_{p}(1).
\]
Therefore,
\begin{align*}
\sup_{\beta\in\mathcal{B}}|\hat{Q_{0}}(\beta)-Q_{0}| & \leq||\hat{\Lambda}_{0}-\Lambda_{0}||\sup_{\beta\in\mathcal{B}}||\hat{g}_{n}(\beta)||^{2}\\
 & \;\;+||\Lambda_{0}||\sup_{\beta\in\mathcal{B}}||\hat{g}_{n}(\beta)-\overline{g}(\beta)||\left(\sup_{\beta\in\mathcal{B}}||\hat{g}_{n}(\beta)||+\sup_{\beta\in\mathcal{B}}||\overline{g}(\beta)||\right)=o_{p}(1).
\end{align*}
Thus $\hat{Q_{0}}$ converges uniformly in probability to $Q_{0}$.
Since $Q_{0}$ has the unique minimizer $\beta^{0}$, $\hat{\beta}^{(1)}\rightarrow\beta^{0}$.
The same argument applies to $\hat{\Lambda^{*}}$ and corresponding
$\hat{Q^{*}}$ and $\hat{\beta}\rightarrow_{p}\beta^{0}$.

Next, we derive the first-order expansion of the sample moment at
$\beta^{0}$ and derive asymptotic normality. Let
\[
\hat{h}_{n}=\hat{\eta}(\beta^{0})-\eta^{0}(\beta^{0}).
\]
By Assumption R(2),
\[
\hat{g}_{n}(\beta^{0})-\frac{1}{n}\sum_{i=1}^{n}g_{i}(\beta^{0})=\frac{1}{n}\sum_{i=1}^{n}D_{i}\left(\hat{h}_{n}\right)+o\left(\frac{1}{\sqrt{n}}\right),
\]
where $D_{i}(\cdot)$ is linear in the nuisance perturbation. Now
we introduce the population linear operator
\[
\mathcal{D}(h)=\E\left[D_{i}(h)\right].
\]
Then,
\[
\sqrt{n}\left\{ \hat{g}_{n}(\beta^{0})-\frac{1}{n}\sum_{i=1}^{n}g_{i}(\beta^{0})\right\} =\frac{1}{\sqrt{n}}\sum_{i=1}^{n}\left(D_{i}\left(\hat{h}_{n}\right)-\mathcal{D}\left(\hat{h}_{n}\right)\right)+\sqrt{n}\mathcal{D}(\hat{h}_{n})+o_{p}(1),
\]
where the first term is $o_{p}(1)$ by Assumption R(3). Moreover,
by Assumption R(1), $\hat{\eta}(\beta^{0})$ is admissible with probability
approaching one. We use $\hat{h}_{n}=\left(\hat{h}_{p,n},\hat{h}_{m,n}\right)$
to represent the propensity-block and imputation block perturbations.
Since $D_{i}(\cdot)$ is linear
\[
\mathcal{D}(\hat{h}_{n})=\mathcal{D}(\hat{h}_{p,n},0)+\mathcal{D}(0,\hat{h}_{m,n}).
\]
By Assumption R(3) and Corollary \ref{lem:=000020neyman-ortho}, which
implies that both derivatives are zero for every admissible perturbation,
we can prove that $\Pr\left[\mathcal{D}(\hat{h}_{n})=0\right]\rightarrow1$
and therefore 
\[
\sqrt{n}||\mathcal{D}(\hat{h}_{n})||=o_{p}(1),
\]
or equivalently,
\[
\sqrt{n}\left(\hat{g}_{n}(\beta^{0})-\frac{1}{n}\sum_{i=1}^{n}g_{i}\right)=o_{p}(1)\Rightarrow\sqrt{n}\hat{g}_{n}(\beta^{0})=\frac{1}{\sqrt{n}}\sum_{i=1}^{n}g_{i}+o_{p}(1).
\]
Under Assumption SMAR, Overlap, and M, by the central limit theorem
applied on the right hand side,
\[
\sqrt{n}\hat{g}_{n}(\beta^{0})\rightarrow_{d}N(0,V).
\]

We now turn to the estimator. We only provide proof for the optimally
estimated $\hat{\beta}$ here, and the same logic applies to all the
second-stage preliminary estimators $\hat{\beta}^{(1)}$. Because
$\hat{\beta}$ minimizes 
\[
\hat{Q^{*}}(\beta)=\hat{g}_{n}(\beta)^{'}\hat{\Lambda^{*}}\hat{g}_{n}(\beta),
\]
and $\beta^{0}\in int(\mathcal{B})$ while $\hat{\beta}\rightarrow_{p}\beta^{0}$,
$\hat{G}_{n}(\hat{\beta})^{'}\hat{\Lambda^{*}}\hat{g}_{n}(\hat{\beta})\rightarrow_{p}0,$
where $\hat{G}_{n}(\beta)=\frac{\partial}{\partial\beta^{'}}\hat{g}_{n}(\beta)$.
Using the integral form of the mean-value expansion,
\[
\hat{g}_{n}(\hat{\beta})=\hat{g}_{n}(\beta^{0})+\overline{G}_{n}(\hat{\beta}-\beta^{0}),
\]
where $\overline{G}_{n}=\int_{0}^{1}\hat{G}_{n}\left(\beta^{0}+t(\hat{\beta}-\beta^{0})\right)dt$.
Substituting this into the first-order condition gives
\[
\hat{G}_{n}(\hat{\beta})^{'}\hat{\Lambda^{*}}\hat{g}_{n}(\beta^{0})+\hat{G}_{n}(\hat{\beta})^{'}\hat{\Lambda^{*}}\overline{G}_{n}(\hat{\beta}-\beta^{0})=0.
\]
Therefore,
\[
\sqrt{n}(\hat{\beta}-\beta^{0})=-\left(\hat{G}_{n}(\hat{\beta})^{'}\hat{\Lambda^{*}}\overline{G}_{n}\right)^{-1}\hat{G}_{n}(\hat{\beta})^{'}\hat{\Lambda^{*}}\sqrt{n}\hat{g}_{n}(\beta^{0}).
\]
We next show that the matrix multiplier converges to the efficient
GMM matrix. Since $\hat{\beta}\rightarrow_{p}\beta^{0}$, under the
regularity assumptions,
\[
||\hat{G}_{n}(\hat{\beta})-G||\leq\sup_{\beta\in\mathcal{N}}||\hat{G}_{n}(\beta)-G(\beta)||+||G(\hat{\beta})-G||\rightarrow_{p}0,
\]
where the first term is $o_{p}(1)$ by assumption and the second term
converges to 0 by continuity of $G(\beta)$ at $\beta^{0}$. Similarly,
\[
||\overline{G}_{n}-G||\leq\sup_{\beta\in\mathcal{N}}||\hat{G}_{n}(\beta)-G(\beta)||+\sup_{t\in[0,1]}||G\left(\beta^{0}+t(\hat{\beta}-\beta^{0})\right)-G||\rightarrow_{p}0.
\]
Together, it follows that
\[
\hat{G}_{n}(\hat{\beta})^{'}\hat{\Lambda^{*}}\overline{G}_{n}\rightarrow_{p}G^{'}V^{-1}G.
\]
Since $G$ has full rank and $V$ is positive definite, $G^{'}V^{-1}G$
is nonsingular. Therefore
\[
\left(\hat{G}_{n}(\hat{\beta})^{'}\hat{\Lambda^{*}}\overline{G}_{n}\right)^{-1}\hat{G}_{n}(\hat{\beta})^{'}\hat{\Lambda^{*}}\rightarrow_{p}\left(G^{'}V^{-1}G\right)^{-1}G^{'}V^{-1},
\]
and
\[
\sqrt{n}(\hat{\beta}-\beta^{0})\rightarrow_{d}N(0,\left(G^{'}V^{-1}G\right)^{-1}).
\]

\begin{flushright}
$\square$
\par\end{flushright}

\subsection{Proof of Theorem \ref{thm:=000020efficiency}}

We prove efficiency following the standard three-step procedure. The
logic and notation follow \citet{chaudhuri2016gmm}, despite a different
data-generating process. We are grateful for their clear and inspiring
framework.

\paragraph{Step 1 }

Continue to denote the observed variables in the dataset as $O\equiv(W,R_{1},R_{2},R_{1}X_{1},R_{2}X_{2})$.
Consider a class of parametric submodels indexed by $\theta$ with
density $f_{\theta}(O)$ with respect to a dominating measure, where
\begin{align*}
f_{\theta}(O) & =\left[p_{\theta}^{11}(W,X_{1})f_{\theta}(X_{1},X_{2}\mid W)\right]^{R_{1}R_{2}}\times\left[p_{\theta}^{10}(W,X_{1})f_{\theta}(X_{1}\mid W)\right]^{R_{1}(1-R_{2})}\\
 & \;\;\times\left[p_{\theta}^{01}(W)f_{\theta}(X_{2}\mid W)\right]^{(1-R_{1})R_{2}}\times\left[p_{\theta}^{00}(W)\right]^{(1-R_{1})(1-R_{2})}f_{\theta}(W),
\end{align*}
with
\begin{align*}
p_{\theta}^{11}(W,X_{1}) & =\Pr\left[R_{1}=1,R_{2}=1\mid W,X\right]=\Pr\left[R_{1}=1,R_{2}=1\mid W,X_{1}\right],\\
p_{\theta}^{10}(W,X_{1}) & =\Pr\left[R_{1}=1,R_{2}=0\mid W,X_{1}\right],\\
p_{\theta}^{01}(W) & =\Pr\left[R_{1}=0,R_{2}=1\mid W,X_{2}\right]=\Pr\left[R_{1}=0,R_{2}=1\mid W\right],\\
p_{\theta}^{00}(W) & =\Pr\left[R_{1}=0,R_{2}=0\mid W\right],
\end{align*}
following from SMAR. The score function is given by
\begin{align*}
S_{\theta}(O) & =s_{\theta}(W)+R_{1}R_{2}s_{\theta}(X_{1},X_{2}\mid W)+R_{1}(1-R_{2})s_{\theta}(X_{1}\mid W)+(1-R_{1})R_{2}s_{\theta}(X_{2}\mid W)\\
 & \;\;+\left(R_{1}R_{2}\frac{\dot{p}{}_{\theta}^{11}(W,X_{1})}{p_{\theta}^{11}(W,X_{1})}+R_{1}(1-R_{2})\frac{\dot{p}_{\theta}^{10}(W,X_{1})}{p_{\theta}^{10}(W,X_{1})}+(1-R_{1})R_{2}\frac{\dot{p}_{\theta}^{01}(W)}{p_{\theta}^{01}(W)}+(1-R_{1})(1-R_{2})\frac{\dot{p}_{\theta}^{00}(W)}{p_{\theta}^{00}(W)}\right),
\end{align*}
where $s_{\theta}(\cdot)\equiv\frac{\partial}{\partial\theta}\operatorname{log}f_{\theta}(\cdot)$
denotes the score function of the submodel, and $\dot{p}_{\theta}^{ij}(\cdot)=\frac{\partial}{\partial\theta}p_{\theta}^{ij}(\cdot)$
for $i,j\in\{0,1\}$.

Define $L_{0}^{2}(F):=\left\{ f:\mathcal{O}\rightarrow\mathbb{R}:\E_{F}[f^{2}]<\infty\text{ and }\E_{F}[f]=0\right\} $
for each probability measure. The tangent set $\mathcal{T}\subset L_{0}^{2}(F)$
consists of all functions of the form

\begin{align*}
h(O) & =R_{1}R_{2}f_{11}(W,X_{1},X_{2})+R_{1}(1-R_{2})f_{10}(W,X_{1})+(1-R_{1})R_{2}f_{01}(W,X_{2})+f_{00}(W)\\
 & \;\;+R_{1}R_{2}\frac{\dot{p}{}_{\theta}^{11}(W,X_{1})}{p_{\theta}^{11}(W,X_{1})}+R_{1}(1-R_{2})\frac{\dot{p}_{\theta}^{10}(W,X_{1})}{p_{\theta}^{10}(W,X_{1})}+(1-R_{1})R_{2}\frac{\dot{p}_{\theta}^{01}(W)}{p_{\theta}^{01}(W)}+(1-R_{1})(1-R_{2})\frac{\dot{p}_{\theta}^{00}(W)}{p_{\theta}^{00}(W)}
\end{align*}
where 
\begin{align*}
f_{11}(W,X_{1},X_{2})\text{\ensuremath{\in L_{0}^{2}\left(F\left(X_{1,}X_{2}|W\right)\right),}}f_{10}(W,X_{1}) & \in L_{0}^{2}\left(F(X_{1}|W)\right)\\
f_{01}(W,X_{2})\text{\ensuremath{\in L_{0}^{2}\left(F\left(X_{2}|W\right)\right),}}f_{00}(W) & \in L_{0}^{2}\left(F(W)\right),
\end{align*}
and the functions $p_{ij}$ and$\dot{p}_{ij}$ satisfy, for all $(W,X_{1})$,
\begin{align*}
\dot{p}_{11}(W,X_{1})+\dot{p}_{10}(W,X_{1})+\dot{p}_{01}(W)+\dot{p}_{00}(W) & =0\\
p_{11}(W,X_{1})+p_{10}(W,X_{1})+p_{01}(W)+p_{00}(W) & =1,\forall(X_{1},W).
\end{align*}

\paragraph*{Step 2}

For the second step, we conjecture the influence function and verify
that it lies in the tangent set by establishing pathwise differentiability
of $\beta^{0}$. By the population moment condition, $A\E\left[g(W,X_{1},X_{2};\text{\ensuremath{\beta^{0}}})\right]=0$
for any $d_{\beta}\times d_{g}$ matrix $A$ that maps a possibly
overidentified system into a just-identified one. 
\begin{align*}
\frac{\partial}{\partial\theta}\beta(\theta_{0})\mid_{\theta=\theta_{0}} & =-(AG)^{-1}A\E\left[g(W,X_{1},X_{2};\beta^{0})\frac{\partial}{\partial\theta}logf_{\theta}(W,X)\right]{}_{\theta=\theta_{0}}\\
 & =-(AG)^{-1}A\E\left[g(W,X_{1},X_{2};\beta^{0})\left(s(W)^{'}+s(X_{1}\mid W)^{'}+s(X_{2}\mid W,X_{1})^{'}\right)\right]_{\theta=\theta_{0}}.
\end{align*}

Then, we conjecture the efficient influence function $\varphi$ as
\[
\varphi(A,W,R_{1}X_{1},R_{2}X_{2},R_{1},R_{2};\beta^{0})=-(AG)^{-1}Ag_{aipw}(O;\beta^{0}).
\]
This function is measurable with respect to $O$ and belongs to $\mathcal{T}$
by the construction of $g_{aipw}$. To verify that it is indeed the
efficient influence function, it suffices to show the pathwise differentiability
condition holds. 
\[
\E\left[\text{\ensuremath{\varphi S_{O}^{'}}}\right]=\E\left[g_{aipw}(O;\beta^{0})\left(s(W)^{'}+s(X_{1}\mid W)^{'}+s(X_{2}\mid W,X_{1})^{'}\right)\right].
\]
Recall from the proof of Theorem \ref{thm:=000020robustness=000020SMAR},
$g_{aipw}$ can be reorganized into
\begin{align*}
g_{aipw}(O;\beta) & =\underbrace{\mu(W;\beta)}_{(1)}+\underbrace{\frac{R_{1}R_{2}}{p_{11}}\left(g(X,W;\beta)-\mu(X_{1},W;\beta)\right)}_{(2)}-\underbrace{(1-p_{1})\frac{R_{1}}{p_{1}}\left(\mu_{21}(X_{1},W;\beta)-\mu_{20}(W;\beta)\right)}_{(3)}\\
 & \;\;-\underbrace{(1-p_{1})\frac{R_{1}R_{2}}{p_{11}}\left(g_{2}\left(X_{2},W;\beta\right)-\mu_{21}(X_{1},W;\beta)\right)}_{(4)}\\
 & \;\;+\underbrace{(1-p_{1})\frac{(1-R_{1})R_{2}}{p_{01}}\left(g_{2}\left(X_{2},W;\beta\right)-\mu_{20}(W;\beta)\right)}_{(5)}.
\end{align*}
We verify this term by term. Decompose $S_{O}$ into data-density
and propensity-score parts:
\[
S_{\theta}(O)=s_{\theta}(W)+R_{1}R_{2}s_{\theta}(X_{1},X_{2}\mid W)+R_{1}(1-R_{2})s_{\theta}(X_{1}\mid W)+(1-R_{1})R_{2}s_{\theta}(X_{2}\mid W)+q(R_{1},R_{2},W,X_{1}),
\]
where
\[
q(R_{1},R_{2},W,X_{1})\equiv R_{1}R_{2}\frac{\dot{p}{}_{\theta}^{11}(W,X_{1})}{p_{\theta}^{11}(W,X_{1})}+R_{1}(1-R_{2})\frac{\dot{p}_{\theta}^{10}(W,X_{1})}{p_{\theta}^{10}(W,X_{1})}+(1-R_{1})R_{2}\frac{\dot{p}_{\theta}^{01}(W)}{p_{\theta}^{01}(W)}+(1-R_{1})(1-R_{2})\frac{\dot{p}_{\theta}^{00}(W)}{p_{\theta}^{00}(W)}.
\]
We first show that $\E\left[g_{aipw}(O;\beta^{0})q(R_{1},R_{2},W,X_{1})\right]=0$.
Proceeding term by term and using the law of iterated expectation
conditional on $(W,X_{1})$,
\[
\E\left[(1)\cdot q(R_{1},R_{2},W,X_{1})\right]=\E\left[\mu(W;\beta)\left(\dot{p}_{11}(W,X_{1})+\dot{p}_{10}(W,X_{1})+\dot{p}_{01}(W)+\dot{p}_{00}(W)\right)\right]=0,
\]
and similarly,
\begin{align*}
\E\left[(2)\cdot q(R_{1},R_{2},W,X_{1})\right] & =\E\left[\frac{R_{1}R_{2}}{p_{11}}\left(g(X,W;\beta)-\mu(X_{1},W;\beta)\right)\frac{\dot{p}{}_{\theta}^{11}(W,X_{1})}{p_{\theta}^{11}(W,X_{1})}\right]=0,\\
\E\left[(4)\cdot q(R_{1},R_{2},W,X_{1})\right] & =\E\left[(1-p_{1})\frac{R_{1}R_{2}}{p_{11}}\left(g_{2}\left(X_{2},W;\beta\right)-\mu_{21}(X_{1},W;\beta)\right)\frac{\dot{p}{}_{\theta}^{11}(W,X_{1})}{p_{\theta}^{11}(W,X_{1})}\right]=0,\\
\E\left[(5)\cdot q(R_{1},R_{2},W,X_{1})\right] & =\E\left[(1-p_{1})\frac{(1-R_{1})R_{2}}{p_{01}}\left(g_{2}\left(X_{2},W;\beta\right)-\mu_{20}(W;\beta)\right)\frac{\dot{p}_{\theta}^{01}(W)}{p_{\theta}^{01}(W)}\right]=0.
\end{align*}
For the remaining term $\E\left[(3)\cdot q(R_{1},R_{2},W,X_{1})\right]$,
note the parts involving $R_{1}$ in $q(R_{1},R_{2},W,X_{1})$ are
\[
\frac{R_{1}-p_{\theta}^{1}(W)}{p_{\theta}^{1}(W)(1-p_{\theta}^{1}(W))}\dot{p}_{\theta}^{1}(W)+R_{1}\frac{R_{2}-\dot{p}_{\theta}^{1|1}(W,X_{1})}{p_{\theta}^{1|1}(W,X_{1})(1-p_{\theta}^{1|1}(W,X_{1}))}\dot{p}_{\theta}^{1|1}(W,X_{1}),
\]
where $p_{\theta}^{1}(W)=\Pr\left(R_{1}=1\mid W\right)$ and $p_{\theta}^{1|1}(W,X_{1})=\Pr\left(R_{2}=1\mid W,X_{1},R_{1}=1\right)$.
Then taking expectations conditional on $W$, 
\[
\E\left[(1-p_{1})\frac{R_{1}}{p_{1}}\left(\mu_{21}(X_{1},W;\beta)-\mu_{20}(W;\beta)\right)\frac{R_{1}-p_{\theta}^{1}(W)}{p_{\theta}^{1}(W)(1-p_{\theta}^{1}(W))}\dot{p}_{\theta}^{1}(W)\right]=0.
\]
Then, on $(W,X_{1})$, we get
\[
\E\left[(1-p_{1})\frac{R_{1}}{p_{1}}\left(\mu_{21}(X_{1},W;\beta)-\mu_{20}(W;\beta)\right)\cdot R_{1}\frac{R_{2}-p_{\theta}^{1|1}(W,X_{1})}{p_{\theta}^{1|1}(W,X_{1})(1-p_{\theta}^{1|1}(W,X_{1}))}\dot{p}_{\theta}^{1|1}(W,X_{1})\right]=0.
\]
Next, we compute
\[
\E\left[g_{aipw}\left(s_{\theta}(W)^{'}+R_{1}R_{2}s_{\theta}(X_{1},X_{2}|W)^{'}+R_{1}(1-R_{2})s_{\theta}(X_{1}|W)^{'}+(1-R_{1})R_{2}s_{\theta}(X_{2}|W)^{'}\right)\right].
\]
Using the original AIPW decomposition, 

\begin{align*}
g_{aipw} & =\underbrace{\frac{R_{1}R_{2}}{p_{11}}g(X,W;\beta)}_{(1)}+\underbrace{\left(1-\frac{R_{1}R_{2}}{p_{11}}\right)\mu(W;\beta)}_{(2)}+\underbrace{\left(\frac{R_{1}}{p_{1}}-\frac{R_{1}R_{2}}{p_{11}}\right)\left(g_{1}(X_{1},W;\beta)-\mu_{1}(W;\beta)\right)}_{(3)}\\
 & \;\;+\underbrace{p_{1}\cdot\left(\frac{R_{1}}{p_{1}}-\frac{R_{1}R_{2}}{p_{11}}\right)\left(\mu_{21}(X_{1},W;\beta)-\mu_{20}(W;\beta)\right)}_{(4)}\\
 & \;\;+\underbrace{(1-p_{1})\cdot\left(\frac{(1-R_{1})R_{2}}{p_{01}}-\frac{R_{1}R_{2}}{p_{11}}\right)\left(g_{2}(X_{2},W;\beta)-\mu_{20}(W;\beta)\right)}_{(5)}.
\end{align*}
We expand the above expectation term-by-term: 
\begin{align*}
 & \E\left[(1)\left(s_{\theta}(W)^{'}+R_{1}R_{2}s_{\theta}(X_{1},X_{2}\mid W)^{'}+R_{1}(1-R_{2})s_{\theta}(X_{1}\mid W)^{'}+(1-R_{1})R_{2}s_{\theta}(X_{2}\mid W)^{'}\right)\right]\\
 & =\E\left[g(X,W;\beta)\left(s_{\theta}(W)^{'}+s_{\theta}(X_{1},X_{2}\mid W)^{'}\right)\right]\\
 & =\E\left[g(X,W;\beta)\left(s_{\theta}(W)^{'}+s_{\theta}(X_{1}\mid W)^{'}+s_{\theta}(X_{2}\mid X_{1},W)^{'}\right)\right].
\end{align*}
For the second term, we calculate

\begin{align*}
 & \E\left[(2)\left(s_{\theta}(W)^{'}+R_{1}R_{2}s_{\theta}(X_{1},X_{2}\mid W)^{'}+R_{1}(1-R_{2})s_{\theta}(X_{1}\mid W)^{'}+(1-R_{1})R_{2}s_{\theta}(X_{2}\mid W)^{'}\right)\right]\\
 & =\E\left[\mu(W;\beta)\left(\left(p_{11}-1\right)s_{\theta}(X_{1},X_{2}\mid W)^{'}+\left(p_{1}-p_{11}\right)s_{\theta}(X_{1}\mid W)^{'}+p_{01}s_{\theta}(X_{2}\mid W)^{'}\right)\right]\\
 & =\E\left[\mu(W;\beta)\left(\left(p_{1}-1\right)s_{\theta}(X_{1}\mid W)^{'}+p_{01}s_{\theta}(X_{2}\mid W)^{'}\right)\right]\\
 & =\E\left[\mu(W;\beta)\left(p_{1}-1\right)E\left[s_{\theta}(X_{1}\mid W)^{'}|W\right]+\mu(W;\beta)p_{01}E\left[s_{\theta}(X_{2}\mid W)^{'}|W\right]\right]=0.
\end{align*}
The second equality above follows from the fact that $\E\left[s_{\theta}(X_{1},X_{2}\mid W)|X_{1},W\right]=s_{\theta}(X_{1}\mid W)$.
For the third term,

\begin{align*}
 & \E\left[(3)\left(s_{\theta}(W)^{'}+R_{1}R_{2}s_{\theta}(X_{1},X_{2}\mid W)^{'}+R_{1}(1-R_{2})s_{\theta}(X_{1}\mid W)^{'}+(1-R_{1})R_{2}s_{\theta}(X_{2}\mid W)^{'}\right)\right]\\
 & =\E\left[\left(g_{1}(X_{1},W;\beta)-\mu_{1}(W;\beta)\right)\left(\left(\frac{p_{11}}{p_{1}}-1\right)s_{\theta}(X_{1},X_{2}\mid W)^{'}+\frac{p_{10}}{p_{1}}s_{\theta}(X_{1}\mid W)^{'}\right)\right]\\
 & =\E\left[\frac{p_{10}}{p_{1}}\left(g_{1}(X_{1},W;\beta)-\mu_{1}(W;\beta)\right)\left(s_{\theta}(X_{1}\mid W)^{'}-s_{\theta}(X_{1},X_{2}\mid W)^{'}\right)\right]\\
 & =-\E\left[\frac{p_{10}}{p_{1}}\left(g_{1}(X_{1},W;\beta)-\mu_{1}(W;\beta)\right)s_{\theta}(X_{2}\mid W,X_{1})^{'}\right]=0.
\end{align*}
The last equality holds because $s_{\theta}(X_{2}\mid W,X_{1})\in L_{0}^{2}\left(F\left(X_{2}\mid W,X_{1}\right)\right)$.
Similarly,
\[
\E\left[(4)\left(s_{\theta}(W)^{'}+R_{1}R_{2}s_{\theta}(X_{1},X_{2}\mid W)^{'}+R_{1}(1-R_{2})s_{\theta}(X_{1}\mid W)^{'}+(1-R_{1})R_{2}s_{\theta}(X_{2}\mid W)^{'}\right)\right]=0.
\]
Finally,
\begin{align*}
 & \E\left[(5)\left(s_{\theta}(W)^{'}+R_{1}R_{2}s_{\theta}(X_{1},X_{2}\mid W)^{'}+R_{1}(1-R_{2})s_{\theta}(X_{1}\mid W)^{'}+(1-R_{1})R_{2}s_{\theta}(X_{2}\mid W)^{'}\right)\right]\\
 & =\E\left[\left(1-p_{1}\right)\left(g_{2}(X_{2},W;\beta)-\mu_{20}(W;\beta)\right)\left(s_{\theta}(X_{2}\mid W)^{'}-s_{\theta}(X_{1},X_{2}\mid W)^{'}\right)\right]\\
 & =-\E\left[\left(1-p_{1}\right)\left(g_{2}(X_{2},W;\beta)-\mu_{20}(W;\beta)\right)s_{\theta}(X_{1}\mid W,X_{2})^{'}\right]=0
\end{align*}
since $s_{\theta}(X_{1}\mid W,X_{2})\in L_{0}^{2}\left(F\left(X_{1}\mid W,X_{2}\right)\right)$.

Note that when $d_{g}>d_{\beta}$, the above characterization of the
tangent space is incomplete, and the influence function must also
satisfy the additional restriction 
\[
\left(I_{d_{g}}-G(AG)^{-1}A\right)\E\left[g_{aipw}(O;\beta)S_{\theta}(O)\right]=0.
\]
This restriction is analogous to that in \citet{barnwell2024efficiency}
and is automatically satisfied when $A$ is chosen optimally (see
Proposition 2 in \citet{barnwell2024efficiency}). In this paper,
we focus on parameters defined by full population moments. Extensions
to subpopulation parameters (e.g., $T$ periods with interest in a
subperiod $[a,b]\subset\{1,2,...,R\}$ and a parameter $\beta_{[a,b]}$
defined by $\E\left[g(O;\beta_{[a,b]})\mid a\leq T\leq b\right]$
are discussed in detail in \citet{barnwell2024efficiency}. 

The above argument implies any regular estimator for $\beta^{0}$
is asymptotically linear with influence function $-(AG)^{-1}Ag_{aipw}(O;\beta^{0})$.

\paragraph*{Step 3}

For any weighting matrix $A$, the asymptotic variance of an estimator
with influence function $-(AG)^{-1}Ag_{aipw}(O;\beta^{0})$ is $(AG)^{-1}AVA^{'}\left((AG)^{-1}\right)^{'}$,
where $V$ is defined in Proposition \ref{prop:=000020var=000020of=000020moment=000020function}.
The optimal choice of $A$ that minimizes this variance is given by
$A^{*}=G^{'}V^{-1}$. With it, the efficiency bound is attained by
the influence function
\[
-(A^{*}G)^{-1}A^{*}g_{aipw}(W,R_{1}X_{1},R_{2}X_{2},R_{1},R_{2};\beta^{0})
\]
and the corresponding asymptotic variance is $\Omega=(G^{'}V^{-1}G)^{-1}$.
\begin{flushright}
$\square$
\par\end{flushright}

\subsection{Proof of Corollary \ref{cor:=000020sieve=000020estimates}.}

We verify the high-level assumptions one by one. We first define $r_{n}=K_{n}^{\kappa+\frac{1}{2}}n^{-\frac{1}{2}}+K_{n}^{-\tau}$
and show $||\hat{e}-e^{0}||_{\infty}=O_{p}\left(r_{n}\right)$.

Let $e_{K_{n}}^{0}$ denote the best approximation to the true primitive
nuisance component $e^{0}$ in the sieve space with dimension $K_{n}$.
Then
\[
||\hat{e}-e^{0}||_{\infty}\leq||\hat{e}-e_{K_{n}}^{0}||_{\infty}+||e_{K_{n}}^{0}-e^{0}||_{\infty}.
\]
The second term is the approximation bias. By \citet{chen2007large},
under the stated H\"{o}lder smoothness and compact support conditions,
standard approximation results for sieves imply
\[
||e_{K_{n}}^{0}-e^{0}||_{\infty}=O\left(K_{n}^{-s_{e}/d_{e}}\right).
\]
The first term is the estimation error. For series and spline sieve
estimators, the standard uniform convergence theory (\citet{newey1997convergence,cattaneo2010efficient})
gives
\[
||\hat{e}-e_{K_{n}}^{0}||_{\infty}=O_{p}\left(\sup_{z\in\mathcal{Z}_{e}}||b_{K_{n},e}(z)||\sqrt{\frac{K_{n}}{n}}\right).
\]
Under the regularity condition (4), 
\[
||\hat{e}-e_{0,K_{n}}||_{\infty}=O_{p}\left(K_{n}^{\kappa+\frac{1}{2}}n^{-\frac{1}{2}}\right).
\]
Therefore,
\[
||\hat{e}-e^{0}||_{\infty}=O_{p}\left(r_{n}\right).
\]
This is the sieve rate used throughout the proof. Then by the definition
of $\eta$, we have
\[
\sup_{\beta\in\mathcal{N}}||\hat{\eta}(\beta)-\eta^{0}(\beta)||=O_{p}(r_{n})
\]
on a neighborhood $\mathcal{N}$ of $\beta^{0}$. Under the regularity
condition (5), we have
\begin{align*}
\nu<\frac{1}{4\kappa+2} & \Rightarrow\nu\left(\kappa+\frac{1}{2}\right)-\frac{1}{2}<-\frac{1}{4},\\
\nu>\frac{1}{4\tau-6\kappa} & \Rightarrow-\nu\tau<-\frac{1}{4},
\end{align*}
and therefore
\[
r_{n}=n^{\nu\left(\kappa+\frac{1}{2}\right)-\frac{1}{2}}+n^{-\nu\tau}=o(n^{-\frac{1}{4}}).
\]
Moreover, $\sqrt{n}r_{n}^{2}=o(1)$, since
\[
\sqrt{n}r_{n}^{2}\lesssim n^{\nu\left(2\kappa+1\right)-\frac{1}{2}}+n^{\frac{1}{2}-2\tau\nu}+n^{\left(\kappa+\frac{1}{2}-\tau\right)\nu},
\]
and all three components are negative under the stated restrictions.

Next we prove the high-level assumptions in R one by one. Admissible
estimators are restricted and we first show linearity and its derivative
representation in R(2). The map 
\[
\eta\rightarrow g_{aipw}(O_{i};\beta^{0},\eta)
\]
is continuously G\^{a}teaux differentiable by construction of $g_{aipw}$.
Therefore, there exists a measurable linear map $h\rightarrow D_{i}(h)$
such that
\[
D_{i}(h)=\frac{d}{dr}g_{aipw}(O_{i};\beta^{0},\eta^{0}(\beta^{0})+rh)|_{r=0}.
\]
In particular, $D_{i}(h)$ is linear in $h$. By the dominated convergence
theorem,
\begin{align*}
\E\left[D_{i}(h_{p},0)\right] & =\frac{d}{dr}\E\left[g_{aipw}\left(O;\beta^{0},(p^{0}+rh_{p},m{}^{0})\right)\right]\vert_{r=0}\\
\E\left[D_{i}(0,h_{m})\right] & =\frac{d}{dr}\E\left[g_{aipw}\left(O;\beta^{0},(p^{0},m^{0}+rh_{m})\right)\right]\vert_{r=0}.
\end{align*}

Then we verify the linearization remainder is $o_{p}(\frac{1}{\sqrt{n}})$.
A first-order Taylor expansion around $\eta^{0}(\beta^{0})$ gives
\[
\hat{g}_{n}(\beta^{0})-\frac{1}{n}\sum_{i=1}^{n}g_{aipw}\left(O_{i};\beta^{0},\eta^{0}(\beta^{0})\right)=\frac{1}{n}\sum_{i=1}^{n}D_{i}(\hat{h}_{n})+R_{n},
\]
where $R_{n}$ is the second-order remainder. Since $g_{aipw}$ admits
a first-order expansion with a quadratic remainder, there exists an
integrable envelope $M_{R}(O)$ such that
\[
||R_{n}||\leq\left(\frac{1}{n}\sum_{i=1}^{n}M_{R}(O_{i})\right)||\hat{h}_{n}||_{\infty}^{2}.
\]
By the sieve rate above, $||\hat{h}_{n}||_{\infty}=O_{p}(r_{n})$
and therefore
\[
||R_{n}||=O_{p}(r_{n}^{2})=o_{p}(\frac{1}{\sqrt{n}})
\]
because $\sqrt{n}r_{n}^{2}=o(1)$. Then we get that
\[
\hat{g}_{n}(\beta^{0})-\frac{1}{n}\sum_{i=1}^{n}g_{aipw}\left(O_{i};\beta^{0},\eta^{0}(\beta^{0})\right)=\frac{1}{n}\sum_{i=1}^{n}D_{i}(\hat{h}_{n})+o_{p}(\frac{1}{\sqrt{n}}).
\]
This completes the verification of R(2).

Now we verify that R(3) holds under the regularity assumptions. This
can be proved term by term. We refer to the decomposition in the following
equation 

\begin{align}
g_{aipw}(\beta) & =\underbrace{\frac{R_{1}R_{2}}{p_{11}}\left(g-\E\left[g\mid W\right]\right)}_{(1)}+\underbrace{\E\left[g\mid W\right]}_{(2)}+\underbrace{\left(\frac{R_{1}}{p_{1}}-\frac{R_{1}R_{2}}{p_{11}}\right)\left(g_{1}-\E\left[g_{1}\mid W\right]\right)}_{(3)}\nonumber \\
 & \;\;+\underbrace{p_{1}\left(\frac{R_{1}}{p_{1}}-\frac{R_{1}R_{2}}{p_{11}}\right)\left(\E\left[g_{2}\mid X_{1},W\right]-\E\left[g_{2}\mid W\right]\right)}_{(4)}\label{eq:=000020decom}\\
 & \;\;+\underbrace{(1-p_{1})(\frac{\left(1-R_{1}\right)R_{2}}{p_{01}}-\frac{R_{1}R_{2}}{p_{11}})\left(g_{2}-\E\left[g_{2}\mid W\right]\right)}_{(5)}.\nonumber 
\end{align}
We proceed with the proof for term (3) in \ref{eq:=000020decom}.
The other terms follow the same logic. Define $\overline{\zeta}_{n}\left(\beta,\hat{p},\hat{\mu}(\beta)\right)=\frac{1}{n}\sum_{i=1}^{n}\hat{\nu}_{i}\hat{\tau}_{i}(\beta),$
where
\begin{align*}
\hat{\nu}_{i} & =\frac{R_{1,i}}{\hat{p}_{1,i}}-\frac{R_{1,i}R_{2,i}}{\hat{p}_{11,i}},\\
\hat{\tau}_{i}(\beta) & =g_{1}(X_{1,i},W_{i};\beta)-\hat{\mu}_{1}(W_{i};\beta).
\end{align*}
Using the same argument as in (32) and (33) in the proof of Proposition
2.3 of \citet{chaudhuri2016gmm}, we can obtain
\[
\sqrt{n}\left(\overline{\zeta}_{n}\left(\beta^{0},\hat{p},\hat{\mu}(\beta^{0})\right)-E\left[\overline{\zeta}_{n}\left(\beta^{0},\hat{p},\hat{\mu}(\beta^{0})\right)\right]\right)=o_{p}(1).
\]
Moreover, Proposition 2.4 of \citet{chaudhuri2016gmm} provides a
stronger local stochastic equicontinuity statement
\[
\sup_{|\beta-\beta^{0}|\leq\delta_{n}}\frac{\sqrt{n}|\overline{\zeta}_{n}\left(\beta,\hat{p},\hat{\mu}(\beta)\right)-\overline{\zeta}_{n}\left(\beta^{0},\hat{p},\hat{\mu}(\beta^{0})\right)|}{1+C\sqrt{n}|\beta-\beta^{0}|}=o_{p}(1),
\]
that is,
\[
\sup_{|\beta-\beta^{0}|\leq\delta_{n}}\frac{\vert\frac{1}{\sqrt{n}}\sum_{i=1}^{n}\hat{\nu}_{i}\left(\hat{\tau}_{i}(\beta)-\hat{\tau}_{i}(\beta^{0})\right)\vert}{1+C\sqrt{n}|\beta-\beta^{0}|}=o_{p}(1)
\]
for every positive sequence $\delta_{n}=o(1)$. 

Lastly, by the sieve rate established at the beginning of the proof,
\[
\sup_{\beta\in\mathcal{N}}||\hat{\eta}(\beta)-\eta^{0}(\beta)||=O_{p}(r_{n})=o_{p}(1).
\]
The above statements verify that Assumption R holds.
\begin{flushright}
$\square$
\par\end{flushright}

\subsection{Proof of Proposition \ref{prop:=000020MAR-equiv}.}

It was shown in \citet{chaudhuri2016gmm} that the following moment
function $g_{CG}$ induces the efficient influence function and attains
the semiparametric efficiency bound:
\begin{align*}
g_{CG} & =\frac{R_{1}R_{2}}{p_{11}}g(X,W;\beta)+\left(1-\frac{R_{1}R_{2}}{p_{11}}\right)\E\left[g(X,W;\beta)\mid W\right]\\
 & \;\;+\frac{p_{10}}{(p_{10}+p_{11})}\left(\frac{R_{1}(1-R_{2})}{p_{10}}-\frac{R_{1}R_{2}}{p_{11}}\right)\left(\E\left[g(X,W;\beta)\mid X_{1},W\right]-\E\left[g(X,W;\beta)\mid W\right]\right)\\
 & \;\;+\frac{p_{01}}{(p_{01}+p_{11})}\left(\frac{(1-R_{1})R_{2}}{p_{01}}-\frac{R_{1}R_{2}}{p_{11}}\right)\left(\E\left[g(X,W;\beta)\mid X_{2},W\right]-\E\left[g(X,W;\beta)\mid W\right]\right).
\end{align*}
Moreover, note that $\frac{p_{10}}{(p_{10}+p_{11})}\left(\frac{R_{1}(1-R_{2})}{p_{10}}-\frac{R_{1}R_{2}}{p_{11}}\right)=\left(\frac{R_{1}}{p_{1}}-\frac{R_{1}R_{2}}{p_{11}}\right)$,
so the part that adjusts for missing $X_{1}$ is exactly the same
between $g_{aipw}$ and $g_{CG}$. Hence, the only difference lies
in how the missingness of $X_{2}$ is corrected. To show differences
induced by different information sets, we use the conditional expectation
notation here instead of $\mu$. Define the difference term as 
\begin{align*}
D(O;\beta) & =\underbrace{\frac{p_{01}}{(p_{01}+p_{11})}\left(\frac{(1-R_{1})R_{2}}{p_{01}}-\frac{R_{1}R_{2}}{p_{11}}\right)\left(\E\left[g(X,W;\beta)\mid X_{2},W\right]-\E\left[g(X,W;\beta)\mid W\right]\right)}_{(1)}\\
 & \;\;-\left(1-p_{1}\right)\left\{ \left(\frac{\left(1-R_{1}\right)R_{2}}{p_{01}}-\frac{R_{1}R_{2}}{p_{11}}\right)\left(g_{2}(X_{2},W;\beta)-E\left[g_{2}(X_{2},W;\beta)\mid W\right]\right)\right.\\
 & \;\;\underbrace{\left.-\left(\frac{R_{1}}{p_{1}}-\frac{R_{1}R_{2}}{p_{11}}\right)\left(\E\left[g_{2}(X_{2},W;\beta)\mid X_{1},W\right]-\E\left[g_{2}(X_{2},W;\beta)\mid W\right]\right)\right\} }_{(2)}.
\end{align*}

Let $\mathcal{T}$ denote the tangent space. A generic score $S\in\mathcal{T}$
decomposes into (a) a ``probability block'' and (b) ``observable
blocks.'' Under MAR, the probability block has the form 
\[
S_{prob}=\sum_{r_{1}r_{2}\in\{0,1\}^{2}}\left(\frac{1(R_{1}=r_{1},R_{2}=r_{2})}{p_{r_{1}r_{2}}(W)}\right)b_{r_{1}r_{2}}(W)
\]
for some measurable functions $b_{r_{1}r_{2}}(W)$ with $\E\left[b_{r_{1}r_{2}}(W)\right]=0$
and finite variance. The observable part can be written as
\[
S_{obs}=s_{\theta}(W)+R_{1}R_{2}s_{\theta}(X_{1},X_{2}\mid W)+R_{1}(1-R_{2})s_{\theta}(X_{1}\mid W)+(1-R_{1})R_{2}s_{\theta}(X_{2}\mid W),
\]
and the full score is $S=S_{prob}+S_{obs}$, with $S\in L_{0}^{2}$.
We first show $D(X_{2},W)$ is orthogonal to the  probability block
at $\beta=\beta^{0}$. As shown in \citet{chaudhuri2016gmm}, $E[(1)S_{prob}]=0$,
and the proof for $E[S_{prob}(2)]=0$ follows that
\[
\E\left[(2)S_{prob}\right]=\E\left[\E\left[(2)\mid R_{1},R_{2},W\right]S_{prob}\right]=0,
\]
since $\E\left[(2)\mid R_{1},R_{2},W\right]=0$ under the MAR assumption.

Next we consider the observable blocks. Under MAR, 
\begin{align*}
 & \E\left[\left(1-p_{1}\right)\left(\frac{\left(1-R_{1}\right)R_{2}}{p_{01}}-\frac{R_{1}R_{2}}{p_{11}}\right)\left(g_{2}(X_{2},W;\beta)-\E\left[g_{2}(X_{2},W;\beta)\mid W\right]\right)S_{\theta}(O)'\right]\\
= & \E\left[\left(1-p_{1}\right)\left(g_{2}(X_{2},W;\beta)-\E\left[g_{2}(X_{2},W;\beta)\mid W\right]\right)\left(s_{\theta}(X_{2}\mid W)-s_{\theta}(X_{1},X_{2}\mid W)\right)^{'}\right]\\
= & -\E\left[\left(1-p_{1}\right)\left(g_{2}(X_{2},W;\beta)-\E\left[g_{2}(X_{2},W;\beta)\mid W\right]\right)s_{\theta}(X_{1}\mid X_{2},W)^{'}\right]=0.\\
 & \E\left[\left(1-p_{1}\right)\left(\frac{R_{1}}{p_{1}}-\frac{R_{1}R_{2}}{p_{11}}\right)\left(\E\left[g_{2}(X_{2},W;\beta)\mid X_{1},W\right]-\E\left[g_{2}(X_{2},W;\beta)\mid W\right]\right)S_{\theta}(O)'\right]\\
= & \E\left[\left(1-p_{1}\right)\left(\E\left[g_{2}(X_{2},W;\beta)\mid X_{1},W\right]-\E\left[g_{2}(X_{2},W;\beta)\mid W\right]\right)\left(s_{\theta}(X_{1}\mid W)-s_{\theta}(X_{1},X_{2}\mid W)\right)^{'}\right]\\
= & -\E\left[\left(1-p_{1}\right)\left(\E\left[g_{2}(X_{2},W;\beta)\mid X_{1},W\right]-\E\left[g_{2}(X_{2},W;\beta)\mid W\right]\right)s_{\theta}(X_{2}\mid X_{1},W)^{'}\right]=0
\end{align*}
using the mean zero properties of $s_{\theta}(X_{2}\mid W,X_{1})\in L_{0}^{2}\left(F\left(X_{2}\mid W,X_{1}\right)\right)$
and $s_{\theta}(X_{1}\mid W,X_{2})\in L_{0}^{2}\left(F\left(X_{1}\mid W,X_{2}\right)\right)$.
A similar argument gives $\E\left[(1)S_{O}^{'}\right]=0$. 

Combining these calculations, we conclude that $\E[D(O;\beta^{0})\cdot S]=0$
for all $S\in\mathcal{T}$. Therefore, $g_{CG}$ and $g_{aipw}$ differ
only by a term that is orthogonal to the tangent space $\mathcal{T}.$
\begin{flushright}
$\square$
\par\end{flushright}

\end{appendix}

\bibliographystyle{plainnat}
\bibliography{240710-draft-new}

\subsection*{Acknowledgement}
The author acknowledges the financial support from
the National Natural Science Foundation of China (72403181, 72595844 and 72573071).

\subsection*{Declaration of generative AI and AI-assisted technologies in the
manuscript preparation process}
During the preparation of this work the author used ChatGPT in order
to polish the writing and check typos. After using this tool/service,
the author reviewed and edited the content as needed and takes full
responsibility for the content of the published article.
\clearpage
\end{CJK}

\end{document}